\documentclass[twocolumn]{aastex62}
\graphicspath{{./}{figures/}}
\shortauthors{Schlafly et al.}
\usepackage{textcomp}
\usepackage{hyperref}
\usepackage{amsmath}
\usepackage{amsfonts}
\newcommand{\degree}{\ensuremath{^\circ}}

\newcommand{\sqdeg}{sq.~deg.}

\newcommand{\code}[1]{\texttt{#1}}
\newcommand{\redrock}{\code{Redrock}}
\newcommand{\fiberassign}{\code{fiberassign}}
\newcommand{\desitarget}{\code{desitarget}}
\newcommand{\quasarnet}{\code{QuasarNET}}

\begin{document}

\title{Survey Operations for the Dark Energy Spectroscopic Instrument}
\shorttitle{Survey Operations for DESI}

\author[0000-0002-3569-7421]{Edward~F.~Schlafly}
\affiliation{Space Telescope Science Institute, 3700 San Martin Dr, Baltimore, MD 21218, USA}

\author[0000-0002-8828-5463]{David~Kirkby}
\affiliation{Department of Physics and Astronomy, University of California, Frederick Reines Hall, Irvine, CA 92697, USA}

\author{David~J.~Schlegel}
\affiliation{Lawrence Berkeley National Laboratory, 1 Cyclotron Road, Berkeley, CA 94720, USA}

\author{Adam~D.~Myers}
\affiliation{Department of Physics and Astronomy, University of Wyoming, Laramie, WY 82071, USA}

\author{Anand~Raichoor}
\affiliation{Lawrence Berkeley National Laboratory, 1 Cyclotron Road, Berkeley, CA 94720, USA}

\author{Kyle~Dawson}
\affiliation{Department of Physics and Astronomy, The University of Utah, 115 South 1400 East, Salt Lake City, UT 84112, USA}

\author{Jessica~Aguilar}
\affiliation{Lawrence Berkeley National Laboratory, 1 Cyclotron Road, Berkeley, CA 94720, USA}

\author{Carlos~Allende~Prieto}
\affiliation{Instituto de Astrof\'{i}sica de Canarias, C/ Vía L\'{a}ctea, s/n, E-38205 La Laguna, Tenerife, Spain}
\affiliation{Universidad de La Laguna, Dept. de Astrof\'{\i}sica, E-38206 La Laguna, Tenerife, Spain}

\author[0000-0003-4162-6619]{Stephen~Bailey}
\affiliation{Lawrence Berkeley National Laboratory, 1 Cyclotron Road, Berkeley, CA 94720, USA}

\author[0000-0001-5537-4710]{Segev~BenZvi}
\affiliation{Department of Physics \& Astronomy, University of Rochester, Bausch and Lomb Hall, Rochester, NY 14627-0171, USA}

\author{Jose~Bermejo-Climent}
\affiliation{Department of Physics \& Astronomy, University of Rochester, Bausch and Lomb Hall, Rochester, NY 14627-0171, USA}

\author{David~Brooks}
\affiliation{Department of Physics \& Astronomy, University College London, Gower Street, London, WC1E 6BT, UK}

\author{Axel~de~la~Macorra}
\affiliation{Instituto de F\'{\i}sica, Universidad Nacional Aut\'{o}noma de M\'{e}xico,  Cd. de M\'{e}xico  C.P. 04510,  M\'{e}xico}

\author[0000-0002-4928-4003]{Arjun~Dey}
\affiliation{NSF's NOIRLab, 950 N. Cherry Ave., Tucson, AZ 85719, USA}

\author{Peter~Doel}
\affiliation{Department of Physics \& Astronomy, University College London, Gower Street, London, WC1E 6BT, UK}

\author[0000-0003-2371-3356]{Kevin~Fanning}
\affiliation{The Ohio State University, Columbus, 43210 OH, USA}

\author[0000-0002-3033-7312]{Andreu~Font-Ribera}
\affiliation{Institut de F\'{i}sica d’Altes Energies (IFAE), The Barcelona Institute of Science and Technology, Campus UAB, 08193 Bellaterra Barcelona, Spain}

\author{Jaime~E.~Forero-Romero}
\affiliation{Departamento de F\'isica, Universidad de los Andes, Cra. 1 No. 18A-10, Edificio Ip, CP 111711, Bogot\'a, Colombia}

\author[0000-0002-9370-8360]{Juan~Garc\'ia-Bellido}
\affiliation{Instituto de F\'{\i}sica Te\'{o}rica (IFT) UAM/CSIC, Universidad Aut\'{o}noma de Madrid, Cantoblanco, E-28049, Madrid, Spain}

\author[0000-0003-3142-233X]{Satya~Gontcho A Gontcho}
\affiliation{Lawrence Berkeley National Laboratory, 1 Cyclotron Road, Berkeley, CA 94720, USA}

\author{Julien~Guy}
\affiliation{Lawrence Berkeley National Laboratory, 1 Cyclotron Road, Berkeley, CA 94720, USA}

\author{ChangHoon~Hahn}
\affiliation{Department of Astrophysical Sciences, Princeton University, Princeton NJ 08544, USA}

\author{Klaus~Honscheid}
\affiliation{Center for Cosmology and AstroParticle Physics, The Ohio State University, 191 West Woodruff Avenue, Columbus, OH 43210, USA}
\affiliation{Department of Physics, The Ohio State University, 191 West Woodruff Avenue, Columbus, OH 43210, USA}
\affiliation{The Ohio State University, Columbus, 43210 OH, USA}

\author[0000-0002-6024-466X]{Mustapha~Ishak}
\affiliation{University of Texas, Dallas, 800 W Campbell Rd, Richardson, TX 75080, USA}

\author{St\'{e}phanie~Juneau}
\affiliation{NSF's NOIRLab, 950 N. Cherry Ave., Tucson, AZ 85719, USA}

\author{Robert~Kehoe}
\affiliation{Department of Physics, Southern Methodist University, 3215 Daniel Avenue, Dallas, TX 75275, USA}

\author[0000-0003-3510-7134]{Theodore~Kisner}
\affiliation{Lawrence Berkeley National Laboratory, 1 Cyclotron Road, Berkeley, CA 94720, USA}

\author[0000-0001-6356-7424]{Anthony~Kremin}
\affiliation{Lawrence Berkeley National Laboratory, 1 Cyclotron Road, Berkeley, CA 94720, USA}

\author[0000-0003-1838-8528]{Martin~Landriau}
\affiliation{Lawrence Berkeley National Laboratory, 1 Cyclotron Road, Berkeley, CA 94720, USA}

\author[0000-0002-1172-0754]{Dustin~A.~Lang}
\affiliation{Perimeter Institute for Theoretical Physics, 31 Caroline St. North, Waterloo, ON N2L 2Y5, Canada}
\affiliation{Department of Physics and Astronomy, University of Waterloo, 200 University Ave W, Waterloo, ON N2L 3G1, Canada}

\author[0000-0003-2999-4873]{James~Lasker}
\affiliation{Department of Physics, Southern Methodist University, 3215 Daniel Avenue, Dallas, TX 75275, USA}

\author[0000-0003-1887-1018]{Michael~E.~Levi}
\affiliation{Lawrence Berkeley National Laboratory, 1 Cyclotron Road, Berkeley, CA 94720, USA}

\author{Christophe~Magneville}
\affiliation{IRFU, CEA, Universit\'{e} Paris-Saclay, F-91191 Gif-sur-Yvette, France}

\author[0000-0003-1543-5405]{Christopher~J.~Manser}
\affiliation{Astrophysics Group, Department of Physics, Imperial College London, Prince Consort Rd, London, SW7 2AZ, UK}
\affiliation{Department of Physics, University of Warwick, Coventry, CV4 7AL, UK}

\author[0000-0002-4279-4182]{Paul~Martini}
\affiliation{Center for Cosmology and AstroParticle Physics, The Ohio State University, 191 West Woodruff Avenue, Columbus, OH 43210, USA}
\affiliation{Department of Astronomy, The Ohio State University, 4055 McPherson Laboratory, 140 W 18th Avenue, Columbus, OH 43210, USA}

\author[0000-0002-1125-7384]{Aaron~M.~Meisner}
\affiliation{NSF's NOIRLab, 950 N. Cherry Ave., Tucson, AZ 85719, USA}

\author{Ramon~Miquel}
\affiliation{Instituci\'{o} Catalana de Recerca i Estudis Avan\c{c}ats, Passeig de Llu\'{\i}s Companys, 23, 08010 Barcelona, Spain}
\affiliation{Institut de F\'{i}sica d’Altes Energies (IFAE), The Barcelona Institute of Science and Technology, Campus UAB, 08193 Bellaterra Barcelona, Spain}

\author[0000-0002-2733-4559]{John~Moustakas}
\affiliation{Department of Physics and Astronomy, Siena College, 515 Loudon Road, Loudonville, NY 12211, USA}

\author[0000-0001-8684-2222]{Jeffrey~A.~Newman}
\affiliation{Department of Physics \& Astronomy and Pittsburgh Particle Physics, Astrophysics, and Cosmology Center (PITT PACC), University of Pittsburgh, 3941 O'Hara Street, Pittsburgh, PA 15260, USA}

\author[0000-0001-6590-8122]{Jundan Nie}
\affiliation{National Astronomical Observatories, Chinese Academy of Sciences, A20 Datun Rd., Chaoyang District, Beijing, 100012, P.R. China}

\author[0000-0003-3188-784X]{Nathalie.~Palanque-Delabrouille}
\affiliation{IRFU, CEA, Universit\'{e} Paris-Saclay, F-91191 Gif-sur-Yvette, France}
\affiliation{Lawrence Berkeley National Laboratory, 1 Cyclotron Road, Berkeley, CA 94720, USA}

\author[0000-0002-0644-5727]{Will~J.~Percival}
\affiliation{Department of Physics and Astronomy, University of Waterloo, 200 University Ave W, Waterloo, ON N2L 3G1, Canada}
\affiliation{Perimeter Institute for Theoretical Physics, 31 Caroline St. North, Waterloo, ON N2L 2Y5, Canada}
\affiliation{Waterloo Centre for Astrophysics, University of Waterloo, 200 University Ave W, Waterloo, ON N2L 3G1, Canada}

\author{Claire~Poppett}
\affiliation{Lawrence Berkeley National Laboratory, 1 Cyclotron Road, Berkeley, CA 94720, USA}
\affiliation{Space Sciences Laboratory, University of California, Berkeley, 7 Gauss Way, Berkeley, CA  94720, USA}
\affiliation{University of California, Berkeley, 110 Sproul Hall \#5800 Berkeley, CA 94720, USA}

\author[0000-0002-6667-7028]{Constance~Rockosi}
\affiliation{Department of Astronomy and Astrophysics, University of California, Santa Cruz, 1156 High St., Santa Cruz, CA 95065, USA}
\affiliation{University of California Observatories, 1156 High St., Sana Cruz, CA 95065, USA}

\author{Ashley~J.~Ross}
\affiliation{Center for Cosmology and AstroParticle Physics, The Ohio State University, 191 West Woodruff Avenue, Columbus, OH 43210, USA}

\author{Graziano~Rossi}
\affiliation{Department of Physics and Astronomy, Sejong University, Seoul, 143-747, Korea}

\author[0000-0003-1704-0781]{Gregory~Tarl\'{e}}
\affiliation{University of Michigan, Ann Arbor, MI 48109, USA}

\author{Benjamin~A.~Weaver}
\affiliation{NSF's NOIRLab, 950 N. Cherry Ave., Tucson, AZ 85719, USA}

\author[0000-0001-5146-8533]{Christophe~Yèche}
\affiliation{IRFU, CEA, Universit\'{e} Paris-Saclay, F-91191 Gif-sur-Yvette, France}

\author[0000-0001-5381-4372]{Rongpu~Zhou}
\affiliation{Lawrence Berkeley National Laboratory, 1 Cyclotron Road, Berkeley, CA 94720, USA}

\collaboration{(DESI Collaboration)}

\begin{abstract}
The Dark Energy Spectroscopic Instrument (DESI) survey is a spectroscopic survey of tens of millions of galaxies at $0 < z < 3.5$ covering 14,000 \sqdeg\ of the sky.  In its first 1.1 years of survey operations, it has observed more than 14 million galaxies and 4 million stars.  We describe the processes that govern DESI's observations of the 15,000 fields composing the survey.  This includes the planning of each night's observations in the afternoon; automatic selection of fields to observe during the night; real-time assessment of field completeness on the basis of observing conditions during each exposure; reduction, redshifting, and quality assurance of each field of targets in the morning following observation; and updates to the list of future targets to observe on the basis of these results.  We also compare the performance of the survey with historical expectations and find good agreement.  Simulations of the weather and of DESI observations using the real field-selection algorithm show good agreement with the actual observations.  After accounting for major unplanned shutdowns, the dark time survey is progressing about 7\% faster than forecast, which is good agreement given approximations made in the simulations.
\end{abstract}




\section{Introduction} 
\label{sec:intro}

The Dark Energy Spectroscopic Instrument (DESI) began a five year survey to measure redshifts of tens of millions of galaxies and quasars on May 14, 2021.  Galaxies and quasars are selected to cover $0 < z < 3.5$ over 14,000 \sqdeg\ of the sky.  The resulting redshifts will be used to measure the expansion history of the universe and the growth of structure to better understand the nature of dark energy \citep{desi:2016a}.

The DESI survey consists of three programs.  The dark program targets luminous red galaxies, emission line galaxies, and quasars, and covers $0.4 < z < 3.5$ \citep{Zhou:2022, Raichoor:2022, Chaussidon:2022}.  Dark program fields are observed whenever conditions are good and represent 90\% of DESI's effective observing time.  The bright program targets a magnitude-limited sample of bright galaxies with $0 < z < 0.4$, as well as Milky Way stars, and is observed when conditions are not good enough to observe dark fields \citep{Hahn:2022, Cooper:2022}.  The combination of the dark program and the bright program are called the ``main survey.''  Finally, a backup program observes bright stars and is only observed when conditions are too poor to observe bright program fields.

These programs consist of a number of ``tiles,'' which are the combination of a location on the sky and an assignment of fibers to locations in the field.  The aim of operations is to observe these fields as efficiently as possible.  Two strategic goals drive many of the choices made in the DESI operations.  First, we intend to observe in a ``depth-first'' mode, where we observe a given part of the sky to completion and never return to it, rather than a ``breadth-first'' mode where observations are spread over the full footprint each year.  Second, we aim to observe $z>2.1$ quasars four times each to improve the signal-to-noise ratio in the Ly-$\alpha$ forest, which enters into the DESI spectral coverage for redshifts $z>2.1$ \citep{desi:2016a}.  This choice means that no observations may overlap a past observation until the $z>2.1$ quasars have been identified, placing pressure on the survey to rapidly and robustly deliver quasar redshifts.  These two goals are in tension with one another---the depth-first goal means that we intend to make overlapping observations quickly to finish parts of the sky, while the goal of identifying $z>2.1$ quasars means that we must complete analysis of observations before we can make overlapping observations.  

\begin{figure*}
    \centering
    \includegraphics[width=\textwidth]{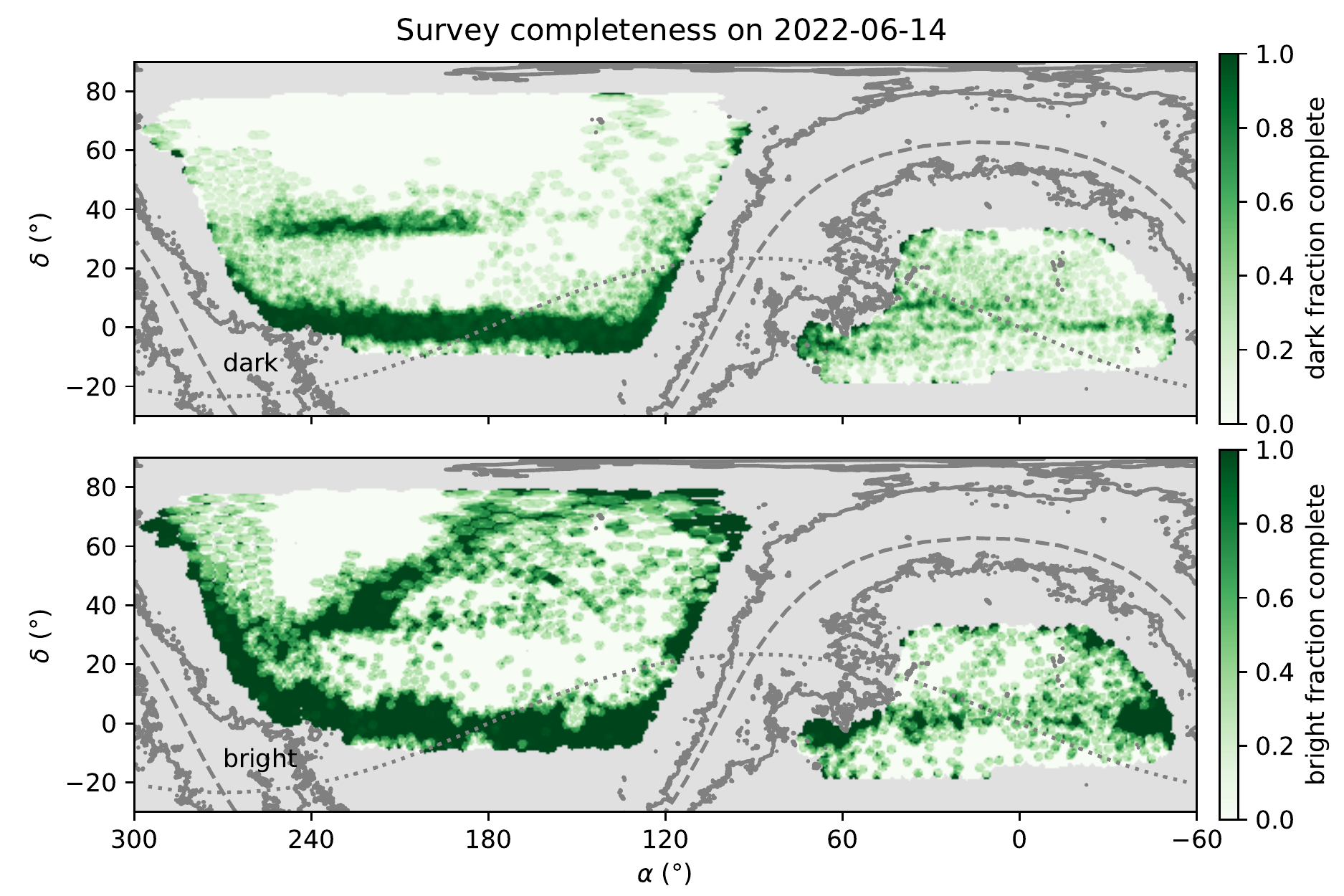}
    \caption{Survey completeness on 2022--06--14, in the dark (top) and bright (bottom) programs.  Green areas are completely finished, while white areas are unfinished.  Areas not included in the footprint are in gray.  Regions with $E(B-V) > 0.3$ are outlined by the solid contours.  The dotted and dashed lines show the ecliptic and Galactic planes.  The survey aims to start observations near $\delta = 0\degree$ and build out.  Notable deviations from that pattern are areas just above $\delta=30\degree$, which are driven by needing to avoid strong winds from the south, and a region $50\degree$ from the ecliptic in the bright program in the north, driven by moon avoidance.}
    \label{fig:progress}
\end{figure*}

Reconciling these goals means bringing together a large number of different processes and analyses on a daily basis to execute the survey.  We focus in this paper on the survey in the time frame from 2021--05--14, the first day of the main survey, to 2022--06--14, when the Contreras wildfire temporarily shut down the survey.  Figure~\ref{fig:progress} shows the area of sky observed by DESI in the dark and bright programs during this period.  We describe the DESI instrument in \textsection\ref{sec:desi}, and elaborate on this broad survey strategy in \textsection\ref{sec:surveystrategy}.  We then describe the different observational and analysis processes that take place on a near-daily basis in order to enable the survey strategy in \textsection\ref{sec:operations}.  The ``merged target list'', which plays a central role in tracking the current state of DESI observations, is described in \textsection\ref{sec:mtldetail}.  The DESI sky footprint is defined in \textsection\ref{sec:footprint}.  The delivered seeing, transparency, sky brightness, and uptime over the first 1.1 years are described in \textsection\ref{sec:performance}.  We detail simulations of the survey in \textsection\ref{sec:surveysim} and compare them with the observed survey performance to date.  Finally, we conclude in \textsection\ref{sec:conclusion}.  The code and data used to produce the tables and figures in this paper are available at \url{https://doi.org/10.5281/zenodo.8010818}.

\section{The Dark Energy Spectroscopic Instrument}
\label{sec:desi}

The Dark Energy Spectroscopic Instrument is a 5000-fiber multi-object spectrograph on the Mayall telescope at Kitt Peak.  The instrument and survey were conceived, designed, and built over a roughly ten year period from 2010--2020 \citep{desi:2013, desi:2016b, desiinstrumentation:2022}.  DESI was designed to measure the expansion history of the universe using the three-dimensional clustering of galaxies and the Lyman-alpha forest over the course of a five-year survey \citep{desi:2016a}.  The instrument collects light from astronomical sources with the 4-m Mayall primary mirror and focuses it through the new corrector onto a 3.2\degree\ diameter focal plane \citep{Miller:2023}.  5000 robotically actuated fibers fill this focal plane \citep{Silber:2023}, piping light through fibers to an array of ten high throughput spectrographs with three channels each spanning the wavelength range 3600--9800 \AA.

The focal plane is divided into ten ``petals,'' nearly identical wedges of the focal plane.  Each petal has 500 positioners, connects to one spectrograph, and contains a guide-focus array imaging camera (GFA).  Four of the petals' GFAs are dedicated to determining the focus of the instrument and deliver out-of-focus images.  The other six deliver in-focus images and are used for guiding, point spread function measurements, and throughput measurements.  The petals are designed to function independently of one another, so that problems with one petal do not affect any other petals.

The main survey will observe millions of stars and galaxies over the course of five years.  Initial results from the survey validation program are now available \citep{desisv, desiedr}.  The primary targets are quasars \citep{Yeche:2020, Chaussidon:2022}, emission line galaxies with $0.6 < z < 1.6$ \citep{Raichoor:2020, Raichoor:2022}, luminous red galaxies with $0.4 < z < 1$ \citep{Zhou:2020, Zhou:2022}, bright galaxies with $z < 0.4$ \citep{RuizMacias:2020, Hahn:2022}, and stars \citep{AllendePrieto:2020, Cooper:2022}.  Targeting catalogs \citep{Myers:2023} for these images were drawn mainly from Data Release 9 of the DESI Legacy Imaging Surveys \citep{Dey:2019}, which included imaging from the Dark Energy Camera on the Blanco telescope \citep{Flaugher:2015}, the 90prime imager on the Bok telescope \citep{Williams:2004, Zou:2017}, and the Mosaic3 imager on the Mayall telescope \citep{Dey:2016}.  Targeting catalogs also incorporated flux and astrometric measurements from Gaia, the Wide-field Infrared Survey Explorer, and the Siena Galaxy Atlas \citep{Gaia:2016, Cutri:2013, fulldepth_neo3, Schlafly:2019, sga}.

Each night, DESI observes roughly twenty tiles containing $\sim$100,000 sources.  By the following morning, the offline pipeline automatically calibrates the resulting exposures, extracts the sources' spectra, subtracts background light, and fits the redshifts of the targets \citep{Guy:2023, redrock:2022}.  The performance of the pipeline was confirmed via a collaboration-wide effort to visually inspect tens of thousands of spectra and their derived redshifts \citep{Lan:2022, Alexander:2022}.

The DESI GFAs and sky monitor provide real-time information on the seeing, transparency, and sky brightness seen by the Mayall \citep{desiinstrumentation:2022, Tie:2020}.  This allows the DESI system to tune the length of exposures to achieve target depths; DESI closes the shutter and reads out the exposure when we have achieved the target signal to noise ratio \citep[][\textsection\ref{subsec:etc}]{Kirkby:2023}.  This process allows us to produce spectra of relatively homogeneous quality even in changing conditions.

\section{Survey Fields}
\label{sec:footprint}

The Dark Energy Survey Instrument Final Design Report calls for a baseline survey of 14,000 \sqdeg\  \citep{desi:2016a}, with a science fiber density of $\sim 3000/\mathrm{deg}^2$ for the dark program and $\sim 700/\mathrm{deg}^2$ for the bright program.  Given the DESI fiber density of $\sim 600/\mathrm{deg}^2$, this corresponds to each region of the sky being covered by five observations for the dark program and one observation for the bright program.  The bright and dark programs nevertheless require more passes to target multiple galaxies within a fiber patrol radius and to obtain reasonable completeness on lower priority main survey programs.  We describe here the specific implementation of these broad requirements for the dark and bright programs.

We define a set of 9929 dark tiles and 5676 bright tiles that cover 14,200 \sqdeg: 9800 \sqdeg\ in the North Galactic Cap and 4400 \sqdeg\ in the South Galactic Cap.  Each tile is a location on the sky that DESI will observe.  These tiles are distributed among several passes where each pass consists of 1,427 non-overlapping tiles.  Approximately 75\% of the footprint can be reached by a DESI fiber in a tile in a particular pass.  The dark program consists of seven such passes, rotated with respect to one another to fill in gaps between the tiles, while the bright program consists of four such passes.  This leads to an average coverage of 5.2 passes for the dark program and 3.2 passes for the bright program.

\begin{figure*}
\plottwo{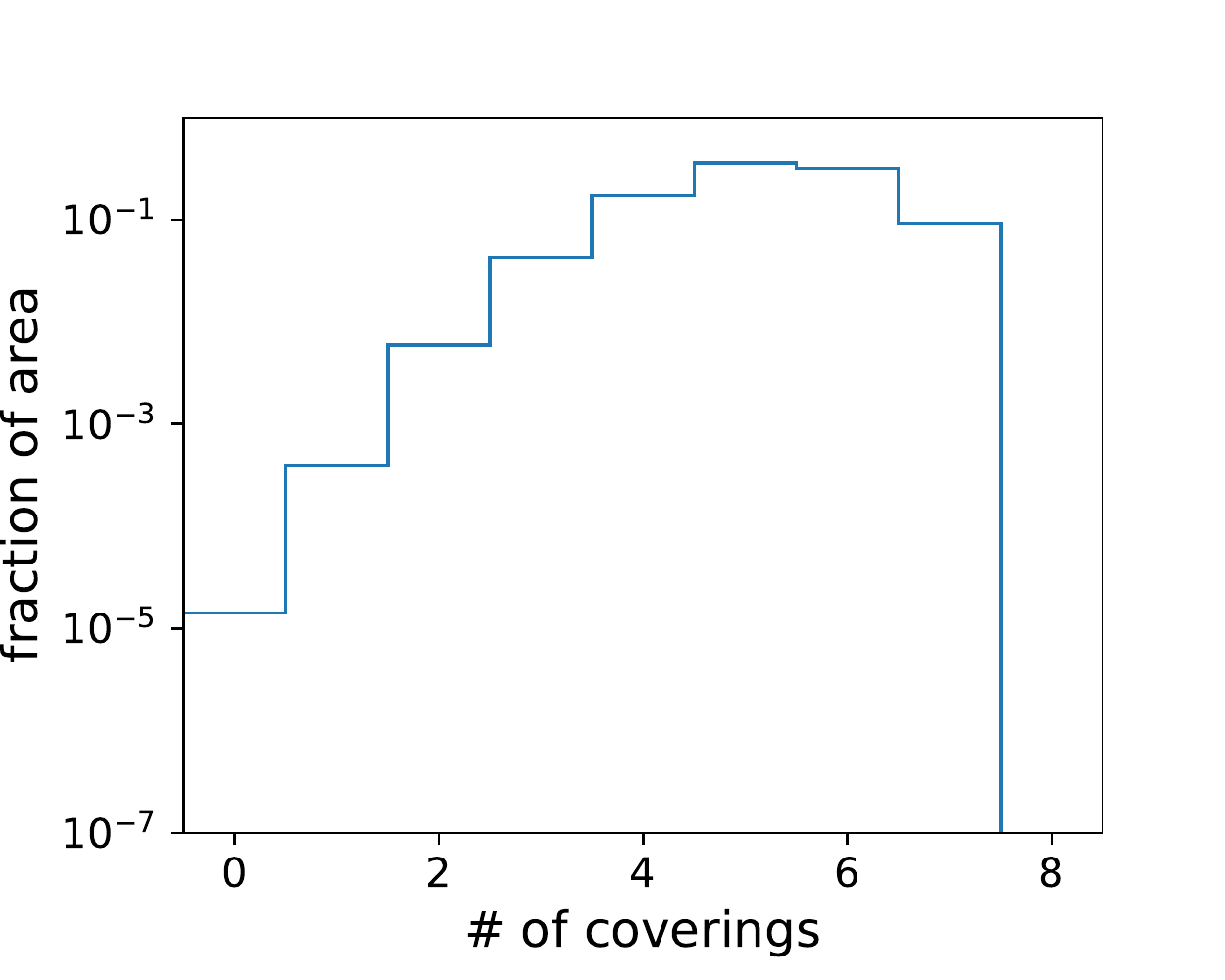}{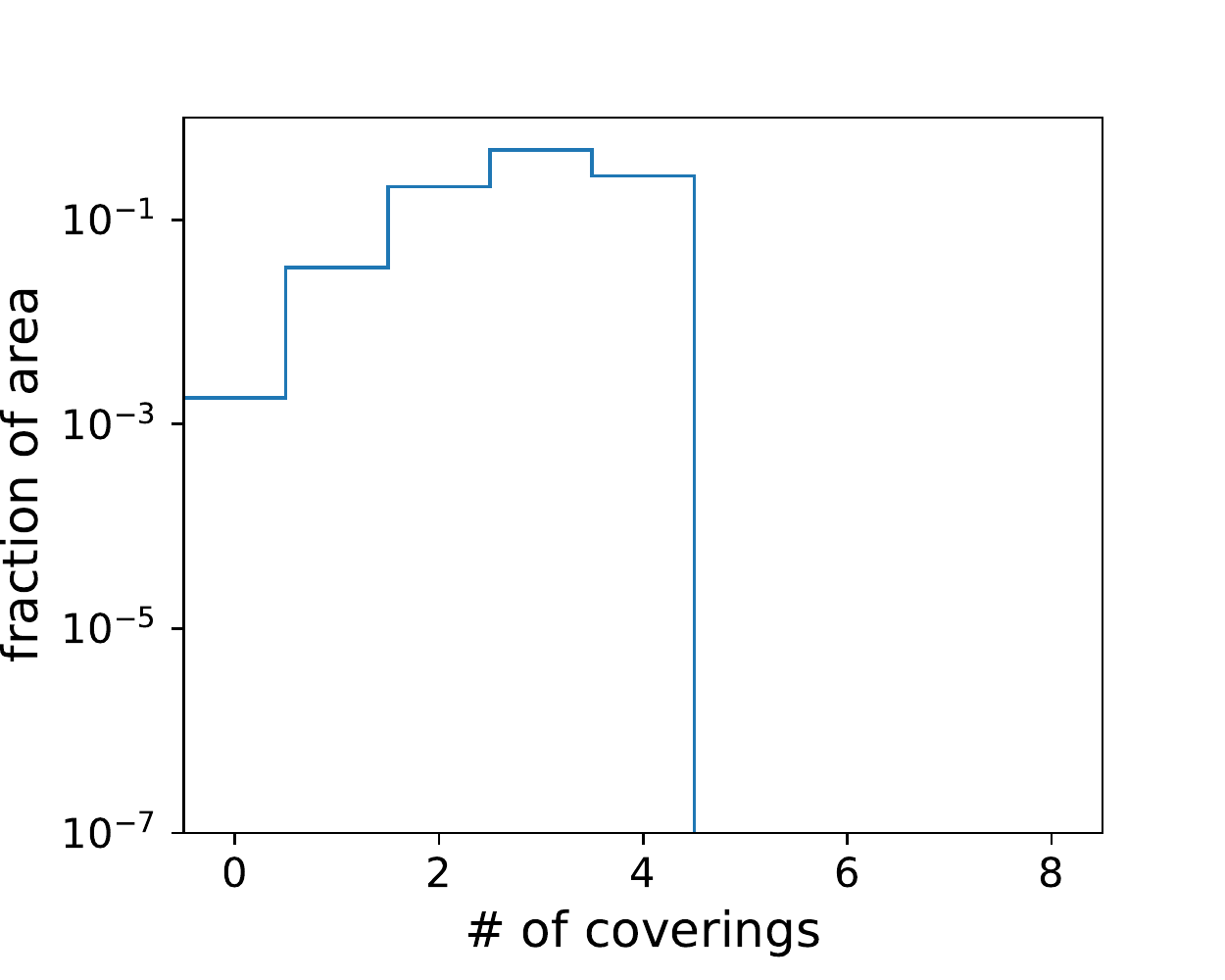}
\caption{The fraction of the sky that is covered by a given number of tiles in the seven-pass dark tiling and the four-pass bright tiling.  On average, a given part of the sky is covered by 5.2 dark tiles and 3.2 bright tiles.
\label{fig:covhist}}
\end{figure*}

\begin{figure*}
\centering
\includegraphics[width=\textwidth]{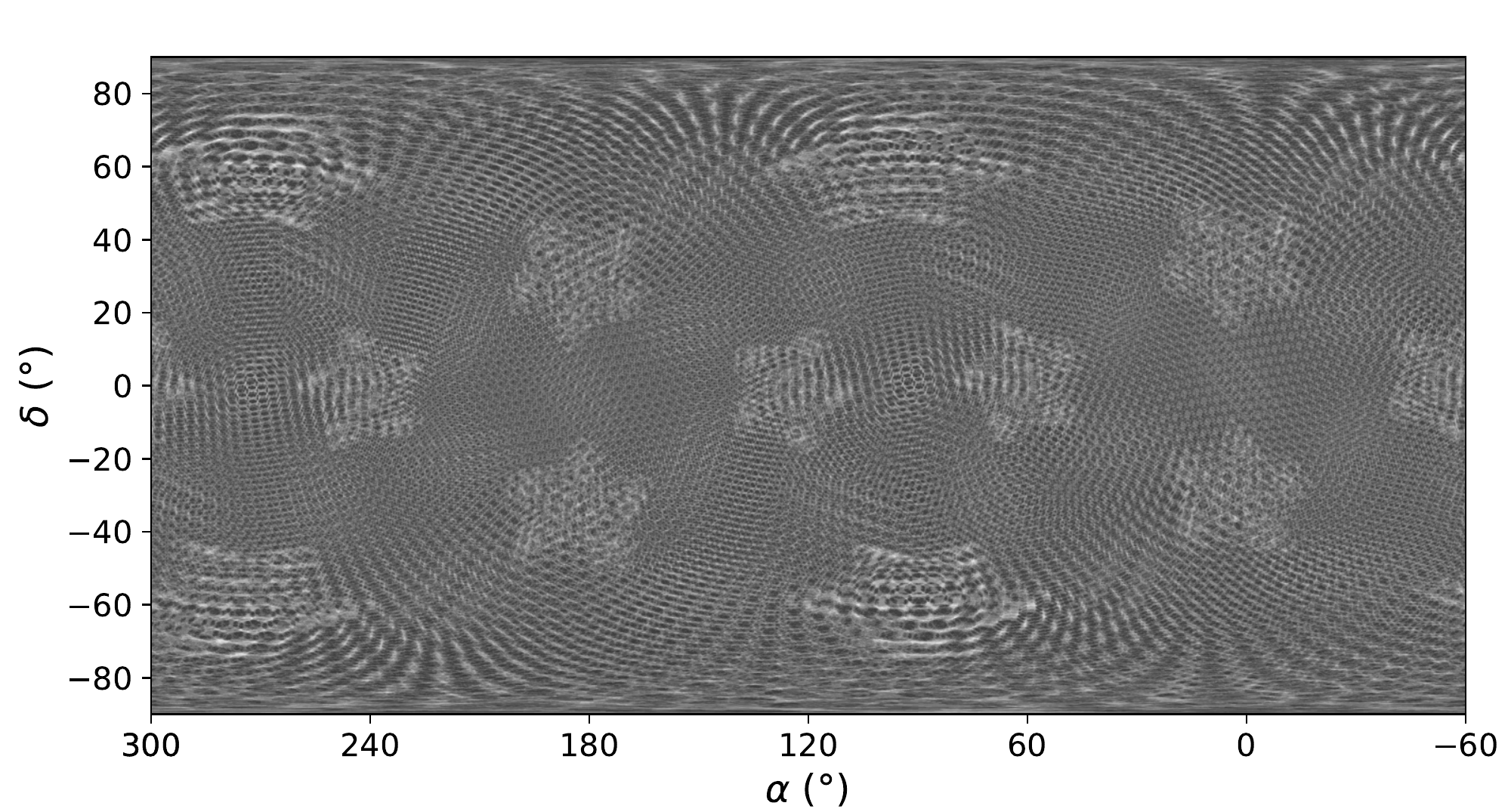}
\caption{The number of exposures that can reach any particular point of the sky, for the seven-pass dark program, were no areas excluded (e.g., due to low Galactic latitude or low declination).  The twelve star-like regions with with slightly lower coverage corresponds to the points of the underlying icosahedral tiling of \citet{Hardin:2000}. 
\label{fig:tilecov}}
\end{figure*}

The pattern of tiles in a single pass is given by the \citet{Hardin:2000} icosahedral tiling with 4112 tile centers distributed over the full sphere.  This tiling matches the size of the DESI focal plane closely and provides a uniform distribution of tiles over the sky.  The fraction of the sky accessible to a given number of tiles for the seven pass dark program and four pass bright program is shown in Figure~\ref{fig:covhist}.  The geometry of the regions of relatively high and low coverage is complicated, and is shown for the seven-pass dark program in Figure~\ref{fig:tilecov}.

The goal of the DESI tile selection was to select a large, contiguous region that could be efficiently observed for extragalactic targets as part of a year-round survey from Kitt Peak.  These objectives imply limits on declination to avoid tiles that are only available at high airmass, and limits on extinction and Galactic latitude to avoid regions where extragalactic targets are both extinguished and more often blended with Milky Way stars.

\begin{figure*}
\includegraphics[width=\textwidth]{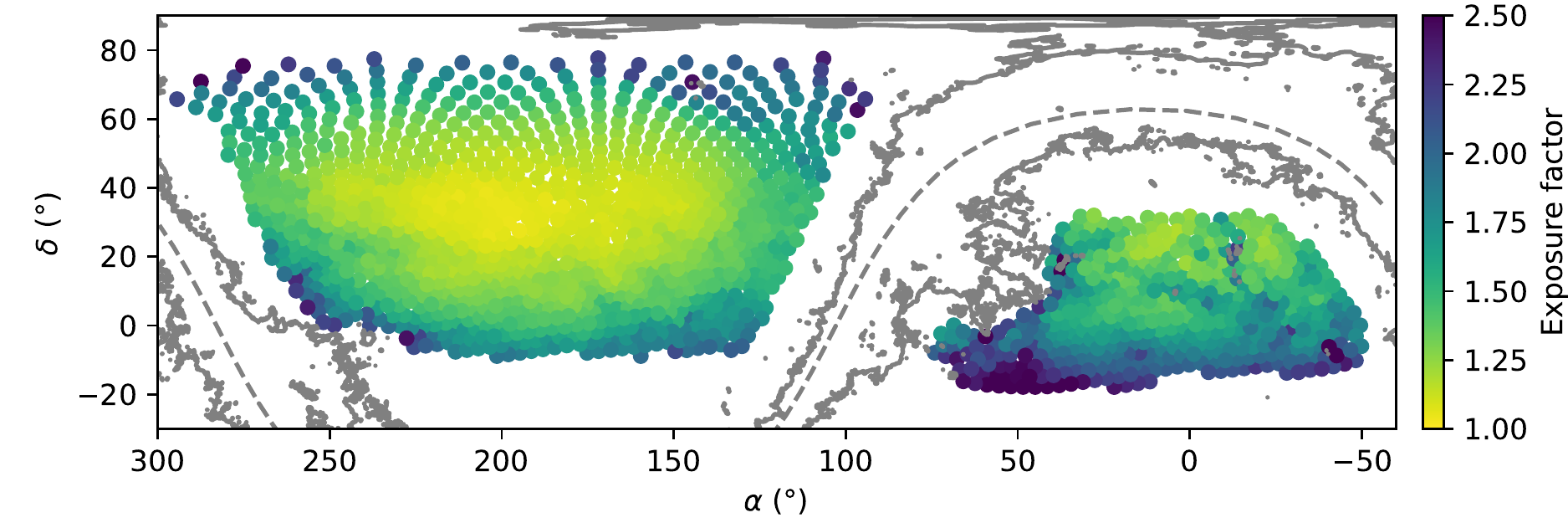}
\caption{The footprint of the DESI survey resulting from the constraints of \textsection\ref{sec:footprint}.  Tiles are colored by the amount of time it would take to reach a fixed intrinsic galaxy depth, relative to observing at zenith in the absence of Galactic extinction.  This is $f_\mathrm{dust} f_\mathrm{airmass}$, from Equations~\ref{eq:dust} and \ref{eq:airmass}.  Airmasses are computed using the design airmasses resulting from the optimization of \textsection\ref{subsec:airmass}.  The Galactic plane is shown as a dotted gray line, and the gray contour shows $E(B-V) = 0.3$ mag.  Tiles in extinguished regions and at the declination bounds of the survey are most expensive, owing to both atmospheric and Galactic extinction.\label{fig:footprint}}
\end{figure*}

We define the footprint as follows:
\begin{enumerate}
\item In the footprint of the DESI Legacy Imaging surveys Data Release 9
\item $-18\degree < \delta < 77.7\degree$
\item $b > 0\degree$ or $\delta < 32.2\degree$
\item $|b| > 22\degree$ for $-90\degree < l < 90\degree$, otherwise $|b| > 20\degree$
\end{enumerate}
These constraints produce the footprint shown in Figure~\ref{fig:footprint}.  Criterion 3 excludes a small portion of the SGC where the Legacy Survey imaging is incomplete.

Though we have imposed no explicit cuts on Galactic extinction, we only target regions of the sky with imaging from the DESI Legacy Imaging Survey.  That survey explicitly avoided high $E(B-V)$ regions, so these regions are naturally avoided in the DESI footprint without need for further adjustment.  Cuts on Galactic latitude do trim the edges of the imaging footprint slightly, however.

The trend in exposure factor with declination in Figure~\ref{fig:footprint} comes from the dependence of survey speed on airmass (\textsection\ref{subsec:speed}).  The SGC is significantly more expensive than the NGC due to a combination of extinction and airmass.  No Legacy Survey imaging was available in the SGC north of $\delta = 32\degree$, though this region would otherwise be favorable for extragalactic studies.  The irregular small-scale variation comes from Galactic extinction.

The sky area within 1.6\degree\ of at least three tile centers for the seven pass dark program is 14,246 \sqdeg.

All main survey tile coordinates are rounded to the nearest 0.001\degree\ for simplicity.

\subsection{Adjustments to tile centers}
\label{subsec:footprintadjustments}
The simple footprint definition of \textsection\ref{sec:footprint} describes our basic footprint selection strategy.  Many tile centers are additionally adjusted to avoid bright stars.

The wide field of view (3.2\degree) of DESI means that bright stars cannot be completely avoided.  However, bright stars are particularly damaging if they fall in a few special parts of the DESI focal plane.

First, it is problematic if a very bright star falls on a GFA.  These can make it challenging to guide the telescope.  Worse, the filter on the GFA reflects light falling outside of the GFA bandpass.  Light from the bright star then ends up adding to a large out-of-focus ghost image covering a substantial portion of the DESI focal plane.  This is avoided by shifting the tile centers to move bright stars off of the GFA filters.  For tiles where a star with Gaia magnitude $G < 6$ lands nears a GFA, we searched for the smallest shift in RA or Dec, in steps of 10 arcseconds, that would put the star at least 25 arcseconds from a GFA.

Second, data from a petal can be rendered useless if a fiber is placed directly on a bright star, saturating large parts of the detector.  This is mostly avoided by re-positioning such fibers (which will never have valid main survey targets) away from bright stars.  But in rare cases a non-functional fiber happens to land on a very bright object.  We adjust tile centers in these cases.  After finding bright stars that land near the current set of non-functional positioners for each tile, we search for a small offset (up to 15 arcseconds) of the tile centers in order to minimize the total star light reaching non-functional positioners.

We periodically compute new offsets for tile centers to account for new or bumped non-functional positioners, but we do not do this on the fly when designing each tile.

\section{Survey Strategy}
\label{sec:surveystrategy}

The goal of DESI is to observe a large, homogeneous, reproducible, and cosmologically interesting set of targets over 14,000 \sqdeg\ of the sky \citep{desi:2016a}.  The survey further aims to operate in a ``depth-first'' fashion where all DESI observations in a particular region are completed before moving on to other parts of the sky.

A critical constraint on the DESI survey strategy is that each DESI observation of a field depends on all earlier, overlapping observations of that field.  This is primarily motivated by the need to identify $z>2.1$ quasars in fields from their initial observations, so that these Ly-$\alpha$ forest tracers can be targeted for repeat observations on subsequent overlapping fields \citep{desi:2016a}.  A secondary motivation is to obtain observations of targets where initial observations failed due to temporary glitches in fiber positioning or in the spectrographs.  This dependence places important constraints on the survey strategy---an observation of a field cannot be made until earlier observations of all overlapping fields have been analyzed.  In particular, no two overlapping fields of either the dark program or the bright program may be observed over the course of a single night.

The DESI survey definition \citep{desi:2016a} provides the basic information about each program, including the targets in each program, the amount of effective exposure time (in essence, signal-to-noise; \textsection\ref{subsec:efftime}) required to observe these targets, and the region of the sky where observations are needed.  Three programs are defined.  First, the dark program, which consists of 9,929 tiles observing luminous red galaxies, emission line galaxies, and quasars from $0.4 < z < 3.5$ (\textsection\ref{sec:footprint}).  Second, the bright program, which consists of 2,657 tiles observing bright galaxies and Milky Way stars (\textsection\ref{sec:footprint}).  Third, the backup program, which consists of brighter Milky Way stars.  Each of these programs have independent target lists that are separately tracked.  The bright and dark programs cover the same region of the high Galactic latitude sky, overlapping spatially; the backup program covers the same area as the bright and dark programs, as well as extending to lower Galactic latitudes.

The dark program is observed whenever conditions are good, and the survey speed for dark tiles is better than $0.4$ (\textsection\ref{subsec:speed}).  When conditions are worse, due to bright skies or poor seeing or transparency, DESI observes the bright program, until the survey speed for bright tiles is worse than $0.08$.  In these poor conditions, DESI observes backup program tiles.  This tiered approach is motivated by placing the brightest targets in the worst conditions, so that systematic uncertainties are limited.  As an added benefit, this approach reduces overheads by placing the exposures needing the shortest effective exposure times in the worst conditions.

The next broad strategic element of the survey is to observe ``depth-first'', completing all DESI observations of a particular region of the sky as soon as possible.  This allows these regions of the sky to be available early for cosmological investigations, and allows many scientific programs to proceed after the first year (albeit over a limited area).  It also minimizes the negative impact of falling behind schedule; we would prefer to end the survey with a complete 13,000 \sqdeg\ survey than an inhomogeneous 14,000 \sqdeg\ survey.  The depth-first goal is implemented in the nightly field selection (\textsection\ref{subsec:nfs}) by preferring tiles near the celestial equator\footnote{A preference for a particular sky region keeps the footprint spatially compact; equatorial fields also enable early science results combining DESI data with other equatorial surveys.}, tiles for which neighboring observations have been made, and tiles which have already been started but for which observations are not yet complete.

The remaining elements of survey planning focus on how we can observe the DESI footprint as efficiently as possible.  This means optimizing the hour angles at which tiles are observed, attempting to observe all tiles as they transit the meridian while reconciling that with the actual distribution of tiles on the sky.  It also means limiting the lengths of the slews between adjacent tiles.

\subsection{Airmass Optimization}
\label{subsec:airmass}

Survey planning assigns each tile an optimal hour angle.  These optimal hour angles need to satisfy two requirements:
\begin{enumerate}
    \item The distribution of local sidereal time (LST) needed to observe all the tiles should match the distribution of local sidereal time expected to be available to the survey.
    \item The total time needed to finish the survey should be as short as possible.
\end{enumerate}
Alternatively, for the dark program, these requirements could be rephrased as asking how to minimize the airmass of the observations subject to the time available to the survey.

The airmass optimization algorithm for DESI is simple.  An initial guess of the assignment of hour angles to tiles is made by matching the LST distribution available to the survey to the right-ascension distribution of the survey's tiles, weighted by the tiles' expected observation times.  The initial assignments of tiles to LSTs is then further optimized through a simulated annealing process to minimize the total amount of time needed to observe the tiles, while maintaining the match between the distribution of LST available to the survey and the distribution of LST needed to observe the tiles.  See appendix \textsection\ref{sec:airmassapp} for more details about the airmass optimization process used in DESI.

Ultimately, the optimization process aims to minimize the expected observation time of the DESI survey.  This is simply the sum of the effective times needed for each tile multiplied by corrections for extinction and airmass.  The extinction correction is given by 
\begin{equation}
\label{eq:dust}
      f_\mathrm{dust} = 10^{2 \times 2.165 \times E(B-V) / 2.5}  \,
\end{equation}
using reddening $E(B-V)$ from \citet{Schlegel:1998} with the calibration of \citet{Schlafly:2011}.  This reddening is taken to be the median SFD reddening over the 3.21\degree\ diameter tile.  Meanwhile the airmass correction is
\begin{equation}
\label{eq:airmass}
    f_\mathrm{airmass} = X^{1.75} \, ,
\end{equation} where $X$ is the airmass of the observation.  This airmass adjustment is an empirical adjustment accounting for lower atmospheric throughput, brighter sky background, and worse seeing at higher airmass.

The DESI airmass optimization scheme is close to optimal for situations when the moon is down.  For the bright time survey when the moon is usually up, determining the optimal observing strategy is much more challenging.  For DESI, this added challenge is ignored and we optimize both the dark and bright programs using the same airmass optimization algorithm---the moon is not included in the optimization process.  The bright program efficiency could be improved by a more advanced optimization process.

The airmass optimization process should be performed periodically as the survey proceeds.  We aim to do this about once a year, but did not update the design hour angles during the first 1.1 years of the survey.

The backup program is not optimized for airmass; we aim to observe all tiles at zero hour angle.  This reflects the fact that completeness and homogeneity are not as important to the backup program as they are to the cosmological programs.

\subsection{Slew Optimization}
\label{subsec:slew}

Long slews reduce the amount of time each night during which DESI can be making science observations.  A number of operations occur when ending one observation and starting a new one \citep{desiinstrumentation:2022}:
\begin{enumerate}
    \item Spectrograph readout
    \item ``Blind'' positioner move
    \item Slewing \& settling
    \item Field acquisition \& guiding
    \item ``Correction'' positioner move
\end{enumerate}
The spectrograph readout and blind positioner move can occur simultaneously with slewing and settling, but the field acquisition and correction move must occur after slewing is complete.  If the slew and settle time exceeds ten seconds, slews begin to increase the overhead between exposures.  Settling time is 8 seconds, and it takes 16 seconds to slew between adjacent DESI fields.  So slewing adds to DESI overheads regardless of slew length.

Nevertheless, even without any explicit slew optimization, slewing would only account for 3.1\% of the open shutter time for the DESI survey, according to survey simulations.  To try to reduce this, we do a simple greedy slew optimization where tiles nearby the current location of the telescope are preferentially observed.  We penalize slews in the declination or negative right ascension directions, but not in the positive right ascension direction, since we do not want to penalize slews that are trying to keep up with the sky rotation.  This simple prescription reduces the slew time to 2.9\% in simulations, and inspection of the resulting slew patterns suggests limited potential for further improvement.

\section{Survey Operations}
\label{sec:operations}

Survey operations broadly refers to the process by which we complete the tiles composing the DESI dark, bright, and backup programs.  Because past exposures inform future exposures, we cannot observe tiles overlapping previously observed ``pending'' tiles until the analysis of those tiles has completed\footnote{Note that the different programs are independent, so a pending bright tile does not block observation of an overlapping dark tile.}.  So the basic operational scheme becomes:
\begin{enumerate}
    \item Each night, observe tiles that do not overlap the footprint of pending tiles.
    \item Each day, analyze observations and incorporate results into the targeting ledger (merged target list or MTL; see \textsection\ref{sec:mtldetail}), clearing pending tiles.
\end{enumerate}
If data reductions are delayed, we may skip step (2), in which case the footprint of pending tiles grows.  We repeat this process until the survey is complete.  The rest of this section details our implementation of this scheme.

The ability to reproduce the particular set of targets that DESI ultimately observes is a key requirement of this process.  We need to be able to simulate the observational process on mock target catalogs in order to account for the effect of the DESI design on the final galaxy redshift catalogs.  Accordingly, we must be capable of reproducing the assignment of every fiber to every target over the course of the survey.  Since these choices depend on the current observational state of the targets and the current health of the instrument, we need to track these quantities through time (see \textsection\ref{sec:mtldetail}, \textsection\ref{subsec:fpsync}).  We record the state of both the targets and the instrument in ledgers.  In these ledgers, each row is time-stamped and changes are made by appending new rows to the ledger indicating the new state of a target or fiber.  Thus, past decisions about the assignment of fibers to targets can be reproduced by reading the ledgers through to the time at which those decisions were made.

\subsection{Daily Observation Overview}
\label{subsec:dailyops}

\begin{figure}
    \centering
    \includegraphics{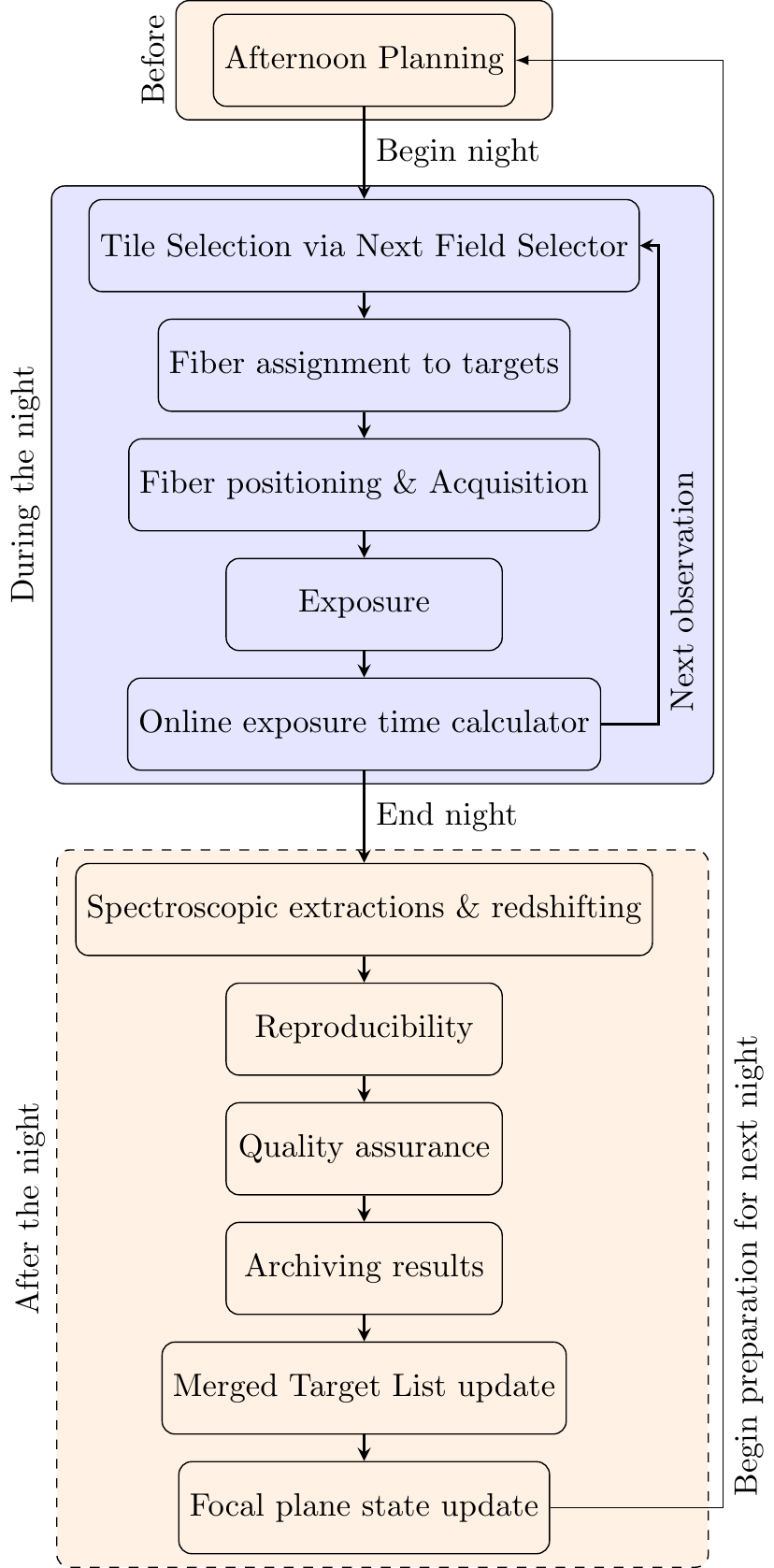}
    \caption{Schematic flow chart of DESI operations steps, running from planning for the night, through each night's observations, through their reduction and updates to the MTL.  Steps in the dashed box are optional and may be skipped temporarily if systems are not available.  See \textsection\ref{subsec:dailyops} for details.}
    \label{fig:flowchart}
\end{figure}

The broad operational model of DESI is specifically implemented in operations in a number of different steps, schematically illustrated in Figure~\ref{fig:flowchart}.  These steps are described in more detail later in this section, and include:
\begin{enumerate}
    \item Afternoon planning identifies completed, pending, and unobserved tiles, and establishes priorities for the night's observations.
    \item The Next Field Selector selects each program and tile to observe during the night.
    \item Targets are assigned to each positioner on the fly immediately before the observation is made.
    \item DESI positions fibers and the spectrograph shutter opens to observe the targets in the field \citep{Silber:2023}.
    \item The Exposure Time Calculator \citep[ETC,][]{Kirkby:2023} computes the effective time obtained on each tile during an observation, determining when an observation is complete.
    \item The spectroscopic pipeline reduces, classifies, and measures redshifts for all targets the following morning \citep{Guy:2023, redrock:2022}.
    \item The reproducibility of the on-the-fly tile design is confirmed by designing the tile a second time outside of operations on the mountain.
    \item Humans perform quality assurance, visually inspecting summary figures and statistics on each tile, and declare tiles either finished or problematic.
    \item Reduced data products for tiles passing quality assurance are archived.
    \item The Merged Target List (MTL) is updated with the new data, updating the observation state and redshift of the observed targets.
    \item The state of the robotic positioners is updated, should any have failed.
    \item The results of the previous nights' observations are available for afternoon planning, and the process repeats for the next night's observations.
\end{enumerate}

Some of these steps need not occur every day.  Pipeline reductions, quality assurance, MTL updates, and focal plane state updates can all be delayed, as illustrated by the dashed box in Figure~\ref{fig:flowchart}.  When MTL updates are delayed, tiles will be left in a ``pending'' state and the survey will be forced to observe new parts of the sky rather than completing the survey in already observed regions.  Delaying focal plane state updates causes only a slightly inefficient assignment of positioners to targets.  In practice, we perform MTL updates roughly weekly in bright time when progress is slow and roughly every other day in good weather in dark time.

The flow chart in Figure~\ref{fig:flowchart} is only intended to be schematic and ignores many details.  For example, the exposure time calculator runs online and is simultaneous with the exposure.  Some visits are split into multiple exposures and do not require new fiber assignment or full positioning \& acquisition loops.  The spectroscopic extractions and redshift determination begin during the night as the data are taken, and so do not strictly follow the separation implied by the flow chart.  Still, Figure~\ref{fig:flowchart} gives a good schematic overview of the DESI daily operation procedure.

\subsection{Effective Time}
\label{subsec:efftime}
The concept of ``effective time'' is important to DESI operations.  We describe effective time briefly here; see \citet{Guy:2023} for more details. Ultimately DESI seeks to measure the fluxes from distant galaxies to a specified accuracy.  Rather than phrasing this accuracy in terms of the flux uncertainty at a particular wavelength, we parameterize it in terms of the amount of time it would take to reach a goal uncertainty in ``nominal'' conditions, defined to be 1.1\arcsec\ seeing, a sky background of 21.07~mag per square arcsecond in the $r$ band, photometric conditions, observations at zenith, through zero Galactic dust reddening.  This ``goal uncertainty'' is weighted over wavelengths and spectral features in order to make it a good proxy for DESI's ability to find a redshift for a galaxy spectrum.  Observations in the dark program aim for 1000~s of effective time, while bright program observations aim for 180~s.

The concept of effective time is made more complicated by the following effects:
\begin{enumerate}
    \item Poisson noise from source flux,
    \item different intrinsic source sizes (e.g., stars versus large galaxies), and
    \item chromatic variation in the sky background and throughput.
\end{enumerate}
The Poisson noise from source flux and the different intrinsic source sizes are challenging because they vary from source to source, making it hard to define the effective time for a tile.  We adopt fiducial  source fluxes and sizes for computing effective times for main survey tiles, which are given in Table~\ref{tab:efftime}.

\begin{deluxetable*}{cccc}
\tablecaption{Source properties used for effective tile effective time\label{tab:efftime}}
\tablehead{\colhead{program} & \colhead{profile} & \colhead{spectrum} & \colhead{source counts}}
\startdata
dark & exponential, $r_\mathrm{half} = 0.45\arcsec$ & LRG spectrum averaged over $0.68 < z < 0.97$ & 0.00 nMgy\\
bright & de Vaucouleurs, $r_\mathrm{half} = 1.5\arcsec$ & BGS spectrum averaged over $0.13 < z < 0.37$ & 1.71 nMgy
\enddata
\tablenotetext{}{See \url{https://www.sdss4.org/dr17/help/glossary/\#nanomaggie} for the definition of nMgy.}
\end{deluxetable*}

Chromatic variation in the system throughput, sky brightness, and detector performance also complicates the notion of effective time.  The goal is to have all tiles reach a nominal depth.  However, for example, when comparing tiles observed through a red, moonless sky with tiles observed through a blue, moony sky, tiles with equal depth in the $r$ band will have different depths in the $g$ and $z$ bands. A simple prescription for this nominal depth would be an average signal-to-noise ratio in a particular range of wavelengths for targets of a given magnitude.  DESI instead adopts a detailed set of weights over all wavelengths that is different for each program, reflecting the spectral lines in the different target classes and their redshift distribution.  See \citet{Guy:2023} for more details.  These more detailed weights are intended to deliver something closer to a uniform redshift success rate for the different key target classes.

Finally, effective time accounts for Galactic extinction.  The ETC \citep[][\textsection\ref{subsec:etc}]{Kirkby:2023} aims to reach a fixed precision in the intrinsic $r$ band flux of target galaxies.  Accordingly, the real time needed to reach a given effective time is increased by Equation~\ref{eq:dust} in the presence of Galactic extinction.

\subsection{Survey Speed}
\label{subsec:speed}
\begin{figure}
\includegraphics[width=\columnwidth]{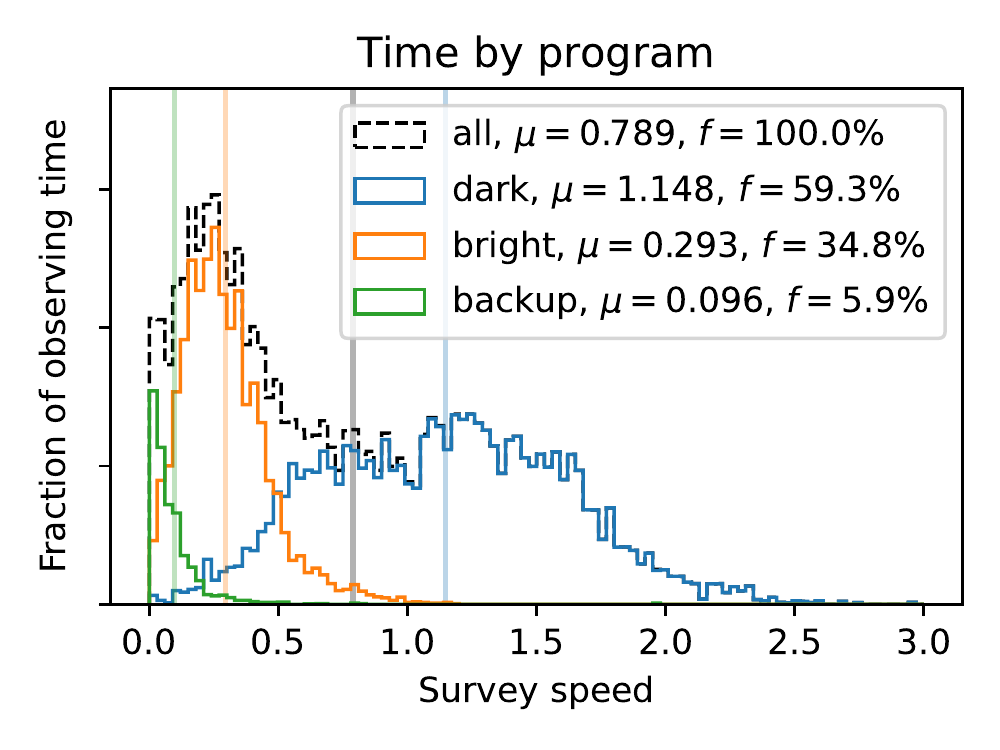}
\caption{The survey speed delivered by the DESI main survey in different programs, as measured by the ETC.  The survey speed describes the rate at which $(S/N)^2$ is accumulated relative to nominal dark conditions, and is highest when the seeing is good and the sky is clear and dark.  The dark program is observed in the best conditions, while the bright and backup programs are observed in progressively worse conditions.  The legend gives the mean speed $\mu$ and the fraction of survey time spent in each program $f$. \label{fig:surveyspeedhist}}
\end{figure}

The concept of survey speed is related to effective time, and is used for a variety of purposes, including the selection of program to observe during the night.  The survey speed is computed using the current seeing, sky background, transparency, and airmass from the Exposure Time Calculator (ETC) (\textsection\ref{subsec:etc}).  The survey speed measures how many effective seconds DESI would be accumulating per second, were DESI observing a tile at zenith and zero dust extinction in the current conditions.  Survey speeds range from zero in clouded-out conditions to $\sim 2.5$ in the best conditions, as shown in Figure~\ref{fig:surveyspeedhist}.  Dark tiles are never observed outside of 15$\degree$ twilight or when survey speed measurements are unavailable, leading to a small number of bright observations in rather good conditions.

The relation between survey speed and seeing depends on the program, since programs observing point sources are more sensitive to seeing than programs observing large galaxies.

The survey speed is adjusted to airmass 1 when observations are made away from zenith following Equation~\ref{eq:airmass}.  This adjustment is intended to account not only for atmospheric extinction, but also for worsened seeing and sky background at lower elevations.  The ETC assesses the survey speed in real time; see \textsection\ref{subsec:etc} for more details.

\subsection{Afternoon Planning}
\label{subsec:ap}

The role of afternoon planning is to determine the current status of survey progress in order to set the base priorities of tiles for the coming night's observations.  Afternoon planning compiles a list of all observed exposures and their associated effective times (\textsection\ref{subsec:efftime}, \textsection\ref{subsec:etc}), and combines these to determine the status of each tile: unobserved, pending, or completed.  This status is used to determine the priority of each tile (\textsection\ref{subsubsec:priorities}), which determines which tiles are observed in the course of the night.  Files describing the configuration of the survey strategy for each night and the state of the survey progress are created.  The Next Field Selector (\textsection\ref{subsec:nfs}) then uses these files in the course of the night's observing.

There are multiple sources for the effective time of each tile.  The authoritative source of this information is the offline pipeline.  Offline pipeline effective times become available in the morning after each night's observations, provided that no issues with the processing or computer systems prevent their computation.  Absent information from the offline pipeline, afternoon planning uses effective times from the ETC (\textsection\ref{subsec:etc}), which are computed on the mountain during each exposure and are always available.

\subsubsection{Tile Priorities}
\label{subsubsec:priorities}

A number of factors contribute to the priority assigned to a tile, which the Next Field Selector uses to select tiles for observation (see \textsection\ref{subsec:nfs}).  Note that these tile priorities are unrelated to the target priorities discussed in \textsection\ref{sec:mtldetail}, which determine which targets get observed within a given tile.  Afternoon planning sets a priority $P$ of each tile for each night according to the following equations:
\begin{align}
\label{eq:tilepriority}
    P & = d s n B \\
    d & = \exp(-|\delta| / 160\degree) \\
    s & = 1 + 0.1 \times \mathtt{is\_started} \\
    n & = 1 + 0.08 \times f_\mathrm{neighbor} \, .
\end{align}
Here $\delta$ is the declination of a tile, \texttt{is\_started} is one if a tile has been started and zero otherwise, and $f_\mathrm{neighbor}$ is the fraction of tiles overlapping this one that have been finished.  The factor $B$ is a rarely used boost factor that can be set to manually change the priority of a tile.

The broad goal of these priorities is to start the survey on the celestial equator and build out ($d$); to finish tiles that have already been started ($s$); and to finish tiles where we already have a number of observations ($n$).  The preference for equatorial tiles keeps the footprint spatially compact and leads to depth-first observations. 
 Starting on the equator also enables early science using cross-correlations with other equatorial surveys.  Finally, it permits follow-up observations of interesting targets from telescopes in both hemispheres.

\subsection{Next Field Selector}
\label{subsec:nfs}

The Next Field Selector (NFS) is responsible for selecting tiles to observe during each night.  Roughly two minutes before each observation is expected to complete, the DESI Instrument Control System (ICS) requests a tile from the NFS.  The NFS selects a program and computes a ``score'' for each tile in that program.  It then chooses the tile with the highest score and designs it on the fly (\textsection\ref{subsec:onthefly}).  The resulting tile is made available to the ICS and is observed.

Program selection is primarily driven by survey speed.  When the survey speed is good, averaging $>0.4$ for the past 20 minutes, dark program tiles are selected.  When the survey speed is poor, $0.08<\mathrm{speed}<0.4$, bright program tiles are selected.  Otherwise, backup tiles are selected.  In addition to this selection, dark tiles are never selected when the sun is within 15\degree\ of the horizon, and bright tiles are not selected when the sun is within 12\degree\ of the horizon.  

The tile scores $S$ used by the NFS are computed as the product of the base tile priority $P$ from afternoon planning (Equation~\ref{eq:tilepriority}), and two additional factors.
\begin{align}
    S & = P e^{-T_\mathrm{slew} / 400~\mathrm{s}} e^{-(H-H_D)^2/2\sigma^2} \\
    \sigma & = (d^2X/dH^2)^{-1/2} / 4
\end{align}
where $T_\mathrm{slew}$ is the estimated time needed to slew to the new tile from the current tile, $H$ is the expected hour angle of the midpoint of the next observation, and $H_D$ is the design hour angle of the tile.  $X(H)$ is the airmass of a tile as a function of its hour angle.

The first factor prefers tiles near the current location of the telescope in order to reduce time spent slewing.  The variable $T_\mathrm{slew}$ is based on the location of the new tile, the current location of the telescope, and the acceleration and cruise speed of the telescope on its hour angle and declination axes.  For the computation of $T_\mathrm{slew}$ in the NFS, we do \emph{not} count time spent slewing in direction of increasing right ascension.  Slews in declination that occur while slewing toward increasing right ascension likewise do not contribute to $T_\mathrm{slew}$.  This is to avoid penalizing the telescope for slewing to keep up with the sky.  We do not, for example, want the telescope to dawdle in one Galactic cap to avoid slewing to the other to keep up with the sky.

The second factor penalizes tiles observed away from their design hour angles.  When observing tiles away from their design hour angles, we prefer to observe high declination tiles to low declination tiles, because the airmass of a low declination tile varies more quickly with hour angle than a high declination tile.  We implement this preference by letting $\sigma$ depend on the second derivative of the airmass with hour angle, evaluated at hour angle zero.  We clip $\sigma$ to between $7.5\degree$ and $15\degree$ to avoid tiles with too-large or too-small observability windows.  This ultimately leads to $\sigma$ taking the value of $15\degree$ above $\delta = 12\degree$, and $\sigma \approx 10\degree$ at the southern boundary of the main footprint.

The NFS also places some constraints that may prevent a tile from being observed.  For example, no tile may be observed within $50\degree$ of the moon, though this limit is occasionally relaxed when the location of the moon in the survey footprint would mean that no tiles were otherwise available.  Similarly, no tiles may be observed within 2\degree\ of a classical planet (one of the first six planets).  Most importantly, no tile may be observed that overlaps a pending tile, as discussed in \textsection\ref{subsec:dailyops}.  Observers may impose additional constraints based on current conditions.  These constraints are most often used to force observation in the north when strong southerly winds would otherwise shake the telescope and degrade the delivered image quality, though they can also be used to chase holes in the clouds.

\subsection{On-the-fly Fiber Assignment}
\label{subsec:onthefly}

Tiles are designed on the fly when requested by the NFS.  This means that we do not know which fibers will be assigned to which targets until minutes before observations begin.  When requested, the \fiberassign\ package \citep{fba} uses the MTL (\textsection\ref{subsec:mtl}, \textsection\ref{sec:mtldetail}) and focal plane state (\textsection\ref{subsec:fpsync}) to determine how best to allocate fibers to targets.  Secondary targets and targets of opportunity are also optionally included.

Tile design takes roughly thirty seconds.  Two minutes are allocated to cover rare cases in dense fields and when latency on the DESI computers is higher than typical.

Because reproducible assignments are critical to the large scale structure analysis of the final redshift catalog, \fiberassign\ inputs are all in the form of ledgers recording the state of the system and targets at any given time.  Moreover, the complete state of the software and input data to \fiberassign\ is logged at run time.  We also recreate each tile designed on-the-fly at the mountain at the National Energy Research Scientific Computing Center (NERSC) on the following day to verify that the same assignments are made (\textsection\ref{subsec:reproducibility}).

On-the-fly assignment is convenient because it allows decisions about which tile should be observed to be made in response to current observing conditions, while also allowing every tile to depend on all of its observed neighbors.  A disadvantage of on-the-fly assignment is that it limits the optimization possibilities of \fiberassign.  In this mode, \fiberassign\ does not know about future observations and cannot adjust its assignment of fibers to targets using that information.

Fiber assignment also needs access to the current state of the DESI focal plane.  A substantial number of DESI positioners ($\sim 700$) cannot be assigned to science targets and are usually left fixed in place.  Small numbers of additional positioners occasionally become non-functional.  In order to optimally assign targets to positioners, \fiberassign\ must avoid assigning functional positioners to locations that would collide with non-functional positioners.  Additionally, we assess whether each non-functional positioner lands on a location which can be used to measure the sky spectrum.  If so, we reduce the number of functional positioners allocated to determining sky.  This has a beneficial impact on survey efficiency, since the number of fibers allocated to sky is nearly 10\% of the total fiber budget, and is similar to the number of non-functional fibers.

\subsection{Exposure Time Calculator}
\label{subsec:etc}

The ETC \citep{Kirkby:2023} is responsible for deciding how long to observe each tile, and how much effective time (\textsection\ref{subsec:efftime}) each tile has accumulated during the night.  It is also responsible for tracking survey speed and deciding when to split long observation sequences into multiple exposures.

The ETC uses measurements of the sky background, seeing, and transparency to perform these tasks.  Sky measurements come from the DESI sky camera, which uses 20 dedicated sky fibers to measure the sky brightness in the $r$ band (two sky fibers on each petal) \citep{desiinstrumentation:2022}.  Seeing and transparency measurements come from the DESI GFAs, which are also used for guiding and focusing the telescope \citep{desiinstrumentation:2022}.  Measurements of the amount of flux entering a fiber relative to nominal---the combination of seeing, throughput, and fiber mis-centering most relevant to the effective time---are computed from GFA frames every eight seconds.

\begin{figure}
\includegraphics[width=\columnwidth]{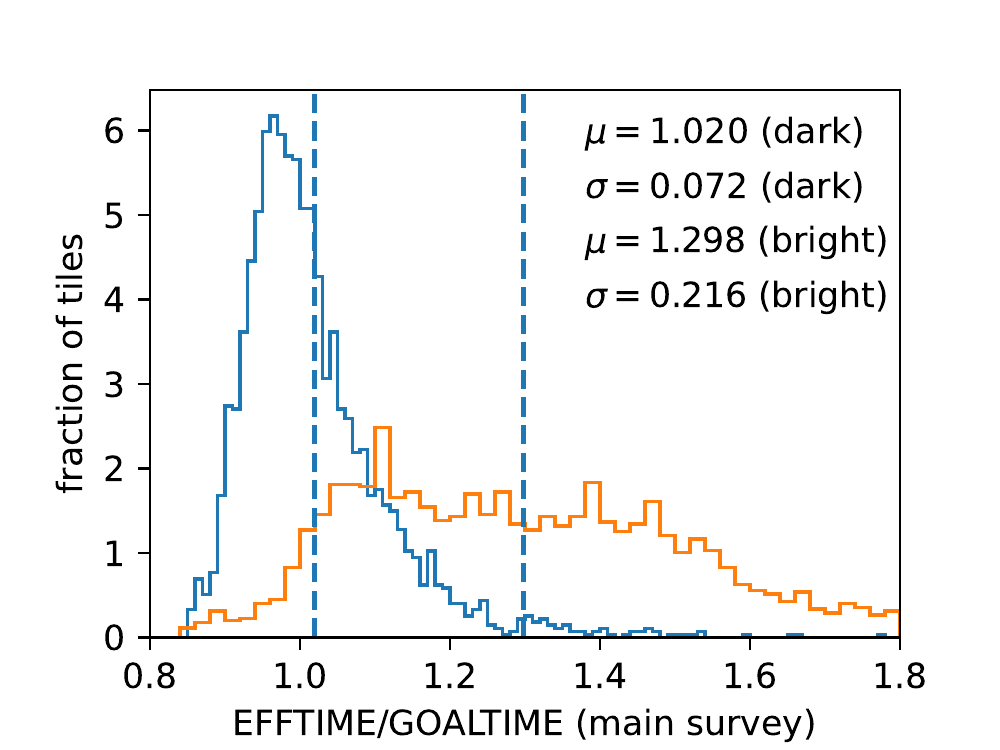}
\caption{Completed dark time tiles have a narrow distribution in EFFTIME around the goal time of 1000 s, with tiles having on average 102\% of their goal effective time, with a standard deviation of 7\% (blue histogram).  This demonstrates that the ETC is able to accurately predict the spectroscopic effective times from the real-time transparency, seeing, and sky brightness measurements.  Bright time tiles have a much broader range of effective time fractions and tend to be observed 30\% longer than necessary (orange histogram).\label{fig:efftimehist}}
\end{figure}

These measurements of the terms contributing to the signal and noise accumulated in the spectrograph are then used to estimate the $(S/N)^2$ obtained in the exposure in real time, which is calibrated to effective time by a single scale factor in each program.  The ETC makes very good predictions for the completeness of dark tiles, leading DESI to have final tile spectroscopic effective times that very closely match their desired goal effective times, as shown in Figure~\ref{fig:efftimehist}.  Bright program tiles show worse agreement due primarily to the varying color of the sky background depending on the phase and location of the moon.  The ETC has access only to the $r$ band sky brightness, while the spectroscopic effective times use the observed brightness of the sky at all wavelengths (\textsection\ref{subsec:efftime}).  Bright program tiles taken in conditions of bright moon tend to be overexposed.

We cap the length of any single exposure to 1800~s for two reasons.  First, long exposures suffer more cosmic ray hits, which wipe out all signal in affected pixels.  By splitting long observations into multiple exposures, a cosmic ray wipes out only the signal in the exposure in which it occurs.  Second, the airmass of a field changes slowly over the course of an exposure.  Splitting long exposures allows us to adjust the atmospheric dispersion corrector for the new location of the field relative to zenith and to reposition the positioners accordingly.  If the ETC determines that an observation is likely to exceed 1800~s, it aims to split it into a series of exposures of equal length.  We cap the amount of time spent on a single tile per night to 90 minutes; if an observation does not reach depth in this time we return to it on a later night.

The required inputs for the ETC are the requested effective time for a tile, the program, and the median Galactic extinction over all targets on each tile.  The requested remaining effective time is provided by the NFS, while the program and extinction are available in the tile files created by \fiberassign.

\subsection{Spectroscopic Pipeline}
\label{subsec:desispec}

The DESI spectroscopic pipeline runs each morning following observations, aiming to complete processing by 10:00 AM Pacific time.  The pipeline carries out a large number of tasks, detailed in \citet{Guy:2023} and \citet{redrock:2022}.  These include:
\begin{enumerate}
    \item processing nightly calibration images (zero second, arc lamp, and flat field exposures),
    \item finding wavelength and two-dimensional line-spread-function solutions for each exposure,
    \item extracting the one-dimensional spectra from the two-dimensional frames after correction for calibration images,
    \item subtracting sky background light,
    \item calibrating spectra to physical units ($10^{-17}$ erg/s/cm$^2$/\AA),
    \item determining redshifts and classifications for each spectrum, and
    \item evaluating the status of each tile and spectrum.
\end{enumerate}
These tasks are all routinely completed within a few hours of the end of the night, for more than $10^5$ fibers on a typical night.

The redshifts are used to update the MTL (\ref{subsec:mtl}), promoting newly detected $z>2.1$ Ly-$\alpha$ quasars to become the highest priority targets on future, overlapping tiles in the dark program.  Other targets are marked with their new redshifts and with flags indicating whether the spectrum is valid or if for some reason the observation should be ignored (e.g., because the positioner did not reach its target location).

\subsection{Fiber Assignment Reproducibility}
\label{subsec:reproducibility}

Galaxy clustering measurements and cosmological analyses of the DESI redshifts depend on being able to reproduce the algorithm by which fibers were assigned to targets.  The on-the-fly assignment of fibers to targets during the night raises concerns that a configuration problem may lead to different assignments when \fiberassign\ (\textsection\ref{subsec:onthefly}) is run on the mountain from when it is run at NERSC.

We reproduce every tile designed over the course of each night at NERSC the following morning to ensure that this does not occur.

\subsection{Quality Assurance}
\label{subsec:qa}

The DESI survey uses the information on each tile to inform later observations of overlapping tiles, via incorporation into the MTL.  The spectroscopic pipeline \citep{Guy:2023, redrock:2022} identifies Ly-$\alpha$ quasars in each observation, so that later tiles can be tasked with reobserving those high-priority targets.  It also identifies which spectra are good, and which spectra are affected by issues with the hardware and should be ignored.

Accordingly, it is important to assess the quality of each observation so that problems with the data are identified before they are incorporated into the MTL.  We make a number of quality assurance (QA) plots for each tile when pipeline reductions of that tile are completed.  These plots include the redshift distribution of the objects on each tile, the redshifts as a function of fiber number, the effective time as a function of location in the focal plane, and the fiber positioning errors as a function of location in the focal plane\footnote{Following fiber positioning, the fiber view camera images the focal plane with the fibers back-lit to identify the final location of the fibers.  The fiber positioning errors shown in QA are the difference between the intended locations and the locations derived from this image.  This is an imperfect proxy; for example, it ignores any systematic errors in the map between true location and location in the fiber view image.  However, it at least highlights any dramatic errors in fiber positioning.}.  The QA also indicates whether the pipeline identified any problems with the tile, like missing standard stars, large reduced-$\chi^2$ values in the sky fibers after sky subtraction, or poor line-spread-function fits.

A member of the operations team reviews the QA for each tile looking for peculiarities.  Most tiles are quickly marked good ($\sim 30$~s per tile).  The remaining more complicated and potentially problematic tiles are marked ``unsure'' and flagged for follow-up investigation.  Examples of such rare cases include tiles with extremely bright stars leading to contamination and sky determination difficulties; cases where small amounts of air leak into the spectrograph, leading to increased glow from the ion pump inside the cryostat and associated enhanced backgrounds; cases where large turbulence in the volume of air between the primary and focal plane causes most positioners to be off target by more than 30~microns RMS; and cases where imperfect sky subtraction in very bright conditions lead to poor redshifts.  Typically exposures affected by these kinds of problems are marked bad and reobserved.

Tiles passing QA are now ready for archiving before inclusion in the MTL (\textsection\ref{subsec:mtl}).

\subsection{Tile Archiving}
\label{subsec:archive}
The daily offline spectroscopic reductions (\textsection\ref{subsec:desispec}) occasionally identify issues in the data or pipeline that need to be addressed before data can be incorporated into the MTL.  In these cases, initial reductions are often deleted and replaced with improved reductions.  For data that eventually enters the MTL, we want to more strictly archive the reductions that were the source of the MTL updates and therefore affect future observations.  Accordingly, once redshift catalogs have been deemed acceptable for incorporation into the MTL, they are copied to a special ``archive'' directory and made read-only.  Updates to the MTL are made only from archived tiles.

\subsection{Merged Target List}
\label{subsec:mtl}

The Merged Target List (MTL) records the current state of each potential DESI target.  Before the survey began, it included entries for each potential target drawn from the imaging surveys, together with the class of that target and its priority.  Following each tile's successful observation and quality assurance check, the archived results of the tile's spectroscopic analysis are used to update the MTL, adjusting the priorities of observed targets.

The most important element of the MTL update is to mark successfully observed objects, so that they may be excluded from future tiles.  The next most important element is to mark newly detected Ly-$\alpha$ quasars as high priority targets which should be observed whenever possible.

These updates are performed by adding new rows to the MTL corresponding to each observed target.  All entries include a timestamp indicating when they were entered into the MTL.  This ledger system enables \fiberassign\ to be run in a reproducible fashion by specifying the latest timestamp in the ledger when \fiberassign\ was run.  Future \fiberassign\ runs can read the ledger through that same timestamp in order to see the same survey state that the original assignment used.  See \textsection\ref{sec:mtldetail} for much more detail about the MTL.

\subsection{Focal Plane State Update}
\label{subsec:fpsync}

The DESI focal plane state describes which positioners are functional, which positioners are not functional, and which regions of the focal plane must be avoided to prevent collisions with non-functional fibers.  The state of the focal plane changes occasionally as positioners malfunction or as positioners are brought back to life.  Malfunctioning positioners are also occasionally moved; this changes the areas of the focal plane which must be avoided.  The operations database at Kitt Peak is the authoritative source of information on the health of each positioner; information from this database must be synced into the state file used by \fiberassign\ in order for fiber assignment to make use of this information.

Like the MTL, the current state of the DESI focal plane is stored in a ledger with timestamps included in every entry.  The state of each positioner at a given point in the history of the instrument can then be obtained by reading the ledger through to that specific time.  We update this ledger via synchronization with the operations database once each day.

Note that the ledger tracking the focal plane state that is used by \fiberassign\ sees only a coarse, daily picture of the state of the positioners.  The online system tracks every move of every positioner and its current state.  When, for example, a positioner fails during a night, \fiberassign\ and the ledger do not see it until the following night.  This means that fiberassign will try to assign targets to non-functional fibers during the night following the failure of a positioner.  The online system then rejects these assignments.  Since at present only roughly one positioner fails per week, there is not much benefit to tracking the focal plane state with better granularity.

Following the focal plane state update, the daily operations loop is ready to repeat.  The MTL and focal plane state have been updated, and afternoon planning (\textsection\ref{subsec:ap}) can prepare for the coming night's observations using the results of the previous night's observations.

\section{Overview of the Merged Target List}
\label{sec:mtldetail}

The Merged Target List (MTL) tracks the observational state of all targets which the DESI survey may observe.  These targets are drawn from a variety of different programs and classes, which may significantly overlap one another, and are denoted by a unique {\tt TARGETID}, as described in \citet{Myers:2023}.  Distinct target classes often need to be treated differently during DESI operations --- for instance $z > 2.1$ quasars ideally need to be observed on 4 overlapping tiles to improve signal-to-noise in the Ly-$\alpha$ forest, whereas emission line galaxies require only a single observation. The main purpose of the DESI MTL code is to enforce a set of decisions for targets that span multiple target classes and so may have competing observational requirements (i.e.\ effectively ``merging'' those targets). In this section, we discuss the form of the various MTL ledgers and the logic used to update them during survey operations.

\subsection{The Initial MTL Ledgers}
\label{subsec:MTLinit}

The MTL software operates on a set of ledgers that contain the minimal information expected to be needed to conduct operational decisions. These ledgers begin with a list of possible targets, which are updated as the survey progresses. Each ledger entry represents a target in a given state at a given time.  Additional entries are added to the end of the ledger when a target's state changes. Crucially, under normal operational procedures, no entries are ever {\em removed} or {\em changed}.  This means that the entire observational history of a target can be recovered by reading a target's ledger entries in order, starting from the initial record.

There are five initial sets of MTL ledgers for the DESI Main Survey: primary dark-time and bright-time ledgers; secondary dark-time and bright-time ledgers; and a set of ledgers for the backup program. Details about how targets are selected for these different programs are available in \citet{Myers:2023}. Structurally, each of these sets of ledgers populates a separate directory and is organized as a set of files split by HEALPixel \citep{Gorski:2005} in the nested scheme at {\tt nside = 32}. This means that each individual ledger covers $\sim3.36\,{\rm deg}^{2}$ of the DESI footprint described in \S\ref{sec:footprint}. Guidelines for creating initial MTL ledgers are included as part of a  tutorial on processing DESI target files that is available on the \desitarget\ GitHub site\footnote{\url{https://github.com/desihub/desitarget/blob/master/doc/nb/how-to-run-target-selection-main-survey.ipynb}}. Details about the data model for, and content of, the MTL ledgers is available as part of the DESI data model\footnote{See \url{https://desidatamodel.readthedocs.io/en/latest/DESI_SURVEYOPS/mtl/index.html}.}.

\subsection{The Initial Observational State}
\label{subsec:MTLpriority}

Each distinct DESI target class has an associated priority and requisite number of observations, which are inherited from the \desitarget\ bitmask ``yaml" files described in \citet{Myers:2023}\footnote{See, e.g., \url{https://github.com/desihub/desitarget/blob/1.1.1/py/desitarget/data/targetmask.yaml} for the DESI Main Survey.}. 
These initial priorities and numbers of observations are stored in the MTL ledgers as {\tt PRIORITY\_INIT} and {\tt NUMOBS\_INIT}. For example, low-priority emission line galaxies ({\tt ELG\_LOP} targets in Table~\ref{table:initprio}) have {\tt PRIORITY\_INIT=3100} and {\tt NUMOBS\_INIT=2}\footnote{As a hedge against potentially needing additional signal-to-noise, {\tt NUMOBS\_INIT} for DESI primary galaxy targets was set to 2 total observations. But, in the DESI Main Survey, the second observation is scheduled at very low priority (see, also \S\ref{subsubsec:genupdates}).}.

A source may be flagged as belonging to multiple target classes. The \texttt{PRIORITY\_INIT} and \texttt{NUMOBS\_INIT} values are set separately for dark-time and bright-time MTL ledgers, using only target classes belonging to the appropriate program. For example, a source could be targeted as a quasar \emph{and} a low-priority emission line galaxy \emph{and} a white dwarf. When constructing the dark-time ledgers, only the quasar and emission line galaxy priorities will be considered; the quasar will ``win'' because \texttt{PRIORITY\_INIT = 3400} (for unobserved quasars) exceeds {\tt PRIORITY\_INIT = 3100} (for unobserved low-priority ELGs). When constructing the bright-time ledgers, only the bright-time white dwarf targeting bit will be considered, because the quasar and emission line galaxy target classes belong to the dark-time program; the white dwarf values will drive the \texttt{PRIORITY\_INIT} and \texttt{NUMOBS\_INIT} settings in the bright-time ledgers. An important principle, here, is that the analysis of the bright-time and dark-time programs are independent.

\begin{deluxetable}{rrl}[t]
\tablecaption{Initial priorities for some DESI target classes\label{table:initprio}}
\tablewidth{0pt}
\tablehead{
\colhead{Target name} & 
\colhead{Priority} & 
\colhead{Notes} 
}
\startdata
\sidehead{\em Dark-time targets}
{\tt QSO} & 3400 & Quasars \\
{\tt LRG} & 3200 & Luminous red galaxies \\
{\tt ELG\_HIP} & 3200 & ELGs at highest priority \\
{\tt ELG\_LOP} & 3100 & ELGs at low priority \\
{\tt ELG\_VLO} & 3000 & ELGs at lowest priority \\
\sidehead{\em Bright-time targets}
{\tt MWS\_WD} & 2988 & White dwarfs \\
{\tt BGS\_BRIGHT} & 2100 & Bright-time galaxies \\
{\tt MWS\_BROAD} & 1400 & General stars \\
\sidehead{\em Rare secondary}
{\tt STRONG\_LENS} & 4000 & Gravitational lenses \\
\sidehead{\em ``Filler'' secondary}
{\tt PSF\_OUT\_DARK} & 90 & Outlier point sources \\
\sidehead{\em Backup targets}
{\tt BACKUP\_GIANT} & 35 & Halo Giants \\
{\tt BACKUP\_FAINT} & 20 & General stars \\
\enddata
\tablenotetext{}{Only a representative subset of target classes is displayed to illustrate the general prioritization schema.}
\end{deluxetable}

\subsubsection{Relative Initial Target Priorities}

The relative initial priorities for targets\footnote{Listed in full in the ``yaml'' files discussed in \S\ref{subsec:MTLpriority}.} are broadly set by a simple underlying philosophy. Lower-density targets are more likely to be swamped by higher-density targets --- so the rarest targets are typically assigned the highest priorities. For example, among dark-time targets, quasars have the highest initial priority, followed by luminous red galaxies and then emission line galaxies. Table~\ref{table:initprio} lists initial priorities for some representative target classes to help illustrate the general schema.

Bright-time targets are always assigned a lower initial priority than dark-time targets. Bright-time galaxies are prioritized over Milky Way targets, regardless of relative density. This ensures that the distribution of Galactic stars is not imprinted on patterns of large-scale structure traced by the bright galaxy program. The sole exception to this scheme is white dwarf targets, which are relatively rare and valuable. Potential white dwarfs are assigned a higher initial priority than all other bright-time targets (but still have a lower initial priority than dark-time targets).

Secondary targets have a range of initial priorities, driven by the intersecting needs of each specific campaign. Secondary targets are generally not allowed to have higher initial priorities than the DESI primary target classes, except for exceedingly rare, high-value targets. Broadly, secondary targets are prioritized by density with very large ``filler'' samples having very low initial priorities. The only targets that have an initial priority lower than ``filler'' secondary classes are targets observed as part of the DESI backup program.

\begin{deluxetable}{ll}[t]
\tablecaption{MTL observational states for DESI targets\label{table:states}}
\tablewidth{0pt}
\tablehead{
\colhead{State} & 
\colhead{Description}
}
\startdata
{\tt UNOBS}          & Unobserved (the {\tt PRIORITY\_INIT} state)        \\
{\tt MORE\_ZWARN}    & Ambigous redshift --- observe more                 \\
{\tt MORE\_ZGOOD}    & Good redshift, but observe more                    \\
{\tt MORE\_MIDZQSO}  & $z\,<\,2.1$ QSO;\,observe\,more\,at\,low\,priority \\
{\tt DONE}           & Enough observations have been obtained             \\
\enddata
\end{deluxetable}

\subsection{Updating the Observational State}
\label{subsec:MTLlogic}

As the DESI survey progresses and redshifts are obtained that reveal the nature of a source, the priority and observational state of a target are updated in the relevant MTL ledger\footnote{These quantities are recorded in the {\tt PRIORITY} and {\tt TARGET\_STATE} columns described in the DESI data model.}. Possible observational states for targets are listed in Table~\ref{table:states}, and each observational state corresponds to a specific numerical priority for a given target class. For example, an unobserved quasar target has a priority of {\tt UNOBS=3400}; a quasar target for which a good redshift is obtained --- $z \geq 2.1$ for a quasar, corresponding to the Ly-$\alpha$ redshift boundary --- has a priority of {\tt MORE\_ZGOOD=3350}; and a quasar target for which observations have been exhausted drops to a priority of {\tt DONE=2}. Setting {\tt MORE\_ZGOOD} $<$ {\tt UNOBS} for quasars ensures that pairs that are closer on the sky than the DESI fiber patrol radius are {\em both} typically observed, because an unobserved quasar has higher priority than one requiring additional observations. The numbers of observations conducted and required for a target are also updated with each acquired redshift, as detailed in \S\ref{subsubsec:genupdates} and \S\ref{subsubsec:qsoupdates}.

\begin{deluxetable*}{ll}[t]
\tablecaption{Flags used to construct the {\tt BAD\_PETALQA} mask\label{table:badpetalqa}}
\tablewidth{0pt}
\tablehead{
\colhead{Flag} & 
\colhead{Description}
}
\startdata
{\tt BADPETALPOS}     & Fraction of fibers with bad positioning ($> 100\,\mu$m) is $> 0.6$ (corresponding to $>300$ fibers on a petal)        \\
{\tt BADPETALSTDSTAR} & Too few standard stars or the rms between stars is too large in the petal (see \S\ref{subsec:zmtl} for more details)\\
{\tt BADREADNOISE}    & Bad readnoise ($> 10$\,electrons/pixel)             \\
\enddata
\begin{center}
\tablenotetext{}{The {\tt BAD\_PETALQA} flag is set if {\em any} bit in this table is set.}
\end{center}
\end{deluxetable*}

\subsubsection{Redshift Information}
\label{subsec:zmtl}

The standard DESI pipeline applies a template-fitting code called \redrock\ \citep{redrock:2022} to derive classifications and redshifts for each target. The MTL code considers redshifts and redshift warnings from \redrock\ when updating the state of a target. These quantities are denoted by {\tt Z} and {\tt ZWARN} in the MTL ledgers, and we adopt this notation below.

The {\tt ZWARN} information from \redrock\ is crucial for the MTL code to determine whether a sufficiently good observation was obtained to update the state of a target. If an observation has a {\tt ZWARN} bit-value of {\tt BAD\_SPECQA}, {\tt BAD\_PETALQA} or {\tt NODATA}\footnote{See, e.g., the {\tt zwarn\_mask} bitmask at \url{https://github.com/desihub/desitarget/blob/2.2.0/py/desitarget/data/targetmask.yaml\#L230-L248}.} set then the observation is considered to not be ``good.'' Such an observation is treated as if it had never been acquired, and the state of the corresponding target is never updated, regardless of the target type. The {\tt NODATA} bit is set by \redrock\ \citep[see][for more details]{redrock:2022}, whereas the {\tt BAD\_SPECQA} and {\tt BAD\_PETALQA} bits --- which we will describe here --- are set as part of the DESI spectroscopic pipeline \citep{Guy:2023}.  Note that a good observation may still correspond to a poor redshift fit, where the most such common redshift failures set the {\tt SMALL\_DELTA\_CHI2} bit for low signal-to-noise spectra.

{\tt BAD\_PETALQA}, which denotes low-quality observations across an entire petal, is flagged when any bit in Table~\ref{table:badpetalqa} is set. Quantitatively, the {\tt BADPETALSTDSTAR} flag listed in Table~\ref{table:badpetalqa}, which denotes a petal that may have insufficient standard stars to extract high-quality spectra, is set when: 
\begin{align}
& N_{\rm good} < 2 \nonumber \\ 
{\rm OR}~~ & N_{\rm good} = 2 ~~{\rm \&}~~ {\rm rms}(R_{\rm flux}) > 0.05 \nonumber \\
{\rm OR}~~ & {\rm rms}(R_{\rm flux}) > 0.2
\end{align}
where $N_{\rm good}$ is the number of good standard stars that the spectroscopic pipeline was able to fit and $R_{\rm flux}$ is the fraction of the expected flux (based on the photometric magnitude) entering the spectrograph.  A standard star is defined as a good fit if
\begin{align}
\chi^2/{\rm dof} < 2 ~~ {\rm \&} ~~ {\rm SNR}({\rm blue})>4 ~~ {\rm \&}  \\ |\Delta(g-r)|<0.1 + 0.2E(B-V) ~.
\end{align}
Here, the ``blue'' region of the spectrum and the $g$- and $r$-camera magnitudes are detailed in \citet{Guy:2023}, and the $E(B-V)$ term allows for some flexibility in the assumed reddening correction.

{\tt BAD\_SPECQA}, which denotes a low-quality spectrum for a single DESI observation, is set when any bit in Table~\ref{table:badpetalqa} {\em or} Table~\ref{table:badspecqa} is flagged. Effective time for a fiber is considered ``too'' low (i.e.\ the {\tt LOWEFFTIME} bit is set) when:
\begin{equation}
{t}_{\rm eff}~10^{2\times2.165~\Delta_{E(B-V)}/2.5} < 0.85\times 0.85 \times {\rm GOALTIME} ~~.
\end{equation}
Here, ${t}_{\rm eff}$ is the effective integration time through the fiber and
\begin{equation*}
\Delta_{E(B-V)} = E(B-V)_{\rm fiber} - {\rm median}(E(B-V)_{\rm tile})
\end{equation*}
accounts for different extinction by Galactic dust through the fiber, as compared to the extinction across the entire tile. The factors of 0.85, which represent the per-tile and per-fiber minimum amount of integration time needed to complete an observation were set by trial-and-error during DESI Survey Validation \citep[e.g.][]{sv}. The quantity on the right-hand side of this inequality ends up being 722 seconds in dark time (${\rm GOALTIME}=1000$\,s) and 130 seconds in bright time (${\rm GOALTIME}=180$\,s), reflecting the effective exposure times listed in \S\ref{subsec:efftime}.

\begin{deluxetable*}{ll}[t]
\tablecaption{Flags used to construct the {\tt BAD\_SPECQA} mask\label{table:badspecqa}}
\tablewidth{0pt}
\tablehead{
\colhead{Flag} & 
\colhead{Description}
}
\startdata
{\tt UNASSIGNED}      & Fiber is not assigned to a known target or sky location  \\
{\tt BROKENFIBER}     & Fiber is broken                                          \\
{\tt MISSINGPOSITION} & Location information is missing for this fiber           \\
{\tt BADPOSITION}     & Fiber was placed $> 100\,\mu$m from the target location  \\
{\tt POORPOSITION}    & Fiber was placed $> 30\,\mu$m from the target location   \\
{\tt LOWEFFTIME}      & Effective time for this fiber is too low (see \S\ref{subsec:zmtl} for more details)       \\
{\tt BADFIBER}        & Fiber is unusable   \\
{\tt BADTRACE}        & Bad trace solution  \\
{\tt BADFLAT}         & Bad fiber flat      \\
{\tt BADARC}          & Bad arc solution    \\
{\tt MANYBADCOL}      & $>10$\% of the pixels covered by this fiber have bad columns \\
{\tt MANYREJECTED}    & $>10$\% of the pixels covered by this fiber were rejected during extraction \\
{\tt BADAMPB}         & Issues with the amplifier readouts of camera B render this fiber unusable  \\
{\tt BADAMPR}         & Issues with the amplifier readouts of camera R render this fiber unusable  \\
{\tt BADAMPZ}         & Issues with the amplifier readouts of camera Z render this fiber unusable  \\
\enddata
\begin{center}
\tablenotetext{}{The {\tt BAD\_SPECQA} flag is set if any bit in this table is set {\em or} if any bit in Table~\ref{table:badpetalqa} is set, although, strictly, {\tt LOWEFFTIME} was not used to inform {\tt BAD\_SPECQA} until April 19, 2022 (see, e.g., \url{https://github.com/desihub/desispec/pull/1722}).}
\end{center}
\end{deluxetable*}

\subsubsection{General Updates}
\label{subsubsec:genupdates}

The MTL uses a ``good'' spectroscopic observation to update the state of most targets via a relatively simple algorithm. The number of required observations (called {\tt NUMOBS\_MORE} in the MTL ledgers) is decremented by one and the number of obtained observations ({\tt NUMOBS}) is incremented by one\footnote{Note that {\tt NUMOBS\_MORE} will equal {\tt NUMOBS\_INIT} for an unobserved target (just as {\tt PRIORITY} will equal {\tt PRIORITY\_INIT}).}. In addition, the {\tt PRIORITY} of a target will be changed to the {\tt MORE\_ZGOOD} or {\tt MORE\_ZWARN} priority if {\tt ZWARN} is zero or non-zero, respectively, for the acquired redshift. As soon as {\tt NUMOBS\_MORE} drops to zero, a target's priority is set to the {\tt DONE} priority discussed in \S\ref{subsec:MTLpriority} (which is a very low value of 2 for all target classes). Similarly, if a target has reached a value equal to the {\tt DONE} priority, then its {\tt NUMOBS\_MORE} value is reduced to zero\footnote{A target can, technically, be observed again once it has reached the {\tt NUMOBS\_MORE=0} state --- such an outcome is simply rendered unlikely because the {\tt DONE} priority is very low.}. Targets for which the {\tt MORE\_ZGOOD} priority is equal to the {\tt DONE} priority will have {\tt NUMOBS\_MORE} drop to zero after their first ${\tt ZWARN}=0$ spectrum is obtained. Similarly, targets for which {\tt MORE\_ZWARN} is equal to {\tt DONE} will no longer be observed after their first observation with ${\tt ZWARN}>0$. The {\tt MORE\_ZGOOD}, {\tt MORE\_ZWARN} and {\tt DONE} priority values are typically identical for both bright-time and dark-time galaxy targets, meaning that such targets are usually only observed once.

\subsubsection{Updates for Quasars}
\label{subsubsec:qsoupdates}

The logic for updating the MTL state is more complex for DESI primary quasar targets and any secondary targets that have {\tt flavor} set to {\tt QSO} in the {\tt scnd\_mask} bitmask\footnote{\url{https://github.com/desihub/desitarget/blob/2.5.0/py/desitarget/data/targetmask.yaml\#L131}.} discussed in \citet{Myers:2023}. In particular, to improve information for Ly-$\alpha$ quasars \citep[e.g.][]{Farr:2020}, the MTL logic incorporates quasar classifications (denoted {\tt IS\_QSO\_QN}) and redshifts (denoted {\tt Z\_QN}) from a line-fitting code called \quasarnet\ \citep{Busca:2018, Green:2023}. 

DESI quasar targets have an initial, unobserved priority of 3400 and are scheduled for 4 total observations. Then, such targets are treated in one of three ways, regardless of whether {\tt ZWARN} indicates the \redrock\ redshift is confident or not:

\begin{itemize}
    \item Quasar targets for which the \redrock\ redshift is ${\tt Z} \geq 2.1$ {\em or} which \quasarnet\ classifies as a definitive high-redshift quasar ({\tt IS\_QSO\_QN==1} {\em and} ${\tt Z\_QN} \geq 2.1$) are denoted ``Ly-$\alpha$'' quasars.
    \item Quasar targets for which the \redrock\ redshift is $1.6 \leq {\tt Z} < 2.1$ {\em and} which \quasarnet\ classifies as a definitive mid-redshift quasar ({\tt IS\_QSO\_QN==1} {\em and} $1.6 \leq {\tt Z\_QN} < 2.1$) are denoted ``mid-z'' quasars.
    \item Otherwise, quasar targets are denoted ``low-z.''
\end{itemize}

Quasars in the ``Ly-$\alpha$'' category have their priority set to {\tt MORE\_ZGOOD} and their {\tt NUMOBS\_MORE} decremented by one. Quasars in the ``mid-z'' category have their priority set to {\tt MORE\_MIDZQSO} and their {\tt NUMOBS\_MORE} decremented by one. Quasars in the ``low-z'' category have their priority set to {\tt MORE\_MIDZQSO} and their {\tt NUMOBS\_MORE} decremented by {\em three}. As with other targets, quasars are observed until their {\tt NUMOBS\_MORE} drops to 0, or below, at which point they are assigned the {\tt DONE} priority and {\tt NUMOBS\_MORE}=0. 

Note that this schema implies that a quasar target can {\em never} reach the {\tt MORE\_ZWARN} state during the DESI Main Survey. Note, also, that ``low-z'' quasars may eventually receive two observations as their {\tt NUMOBS\_MORE} will only drop to {\em one} after their first acquisition. The second observation, however, will be scheduled at a priority ({\tt MORE\_MIDZQSO}) that exceeds only the lowest-priority, highest-density DESI ``filler'' targets. This choice reflects the low density and relatively high scientific value of even $z < 1.6$ and ambiguously classified quasars. 

\subsubsection{Special Cases}
\label{subsubsec:specupdates}

There are two special cases that inform how the MTL ledgers are updated. First, any target that becomes a quasar in the ``Ly-$\alpha$'' category is {\em locked into} that state until it reaches {\tt NUMOBS\_MORE} of 0 and the {\tt DONE} priority. This provides some insurance in the case of genuine $z \geq 2.1$ quasars having a flawed observation or fluctuating in redshift around $z = 2.1$ due to noise. Second, only primary programs are allowed to determine the state in the primary ledgers {\em except} in the case of primary targets that are either for calibration or are only in the Milky Way Survey (MWS) program. Such primary targets {\em are} allowed to be updated by secondary target classes that have {\tt updatemws} set to {\tt True} in the {\tt scnd\_mask} bitmask discussed in \citet{Myers:2023}. This allows the MWS \citep[see][]{Cooper:2022} to better prioritize highly desirable secondary target classes for Galactic science without impacting primary analyses of extragalactic large-scale structure.

\subsubsection{Reprocessing the MTL Ledgers}

Beyond the routine MTL updates discussed in \S\ref{subsubsec:genupdates}, \S\ref{subsubsec:qsoupdates} and \S\ref{subsubsec:specupdates} the MTL ledgers can be fully {\em reprocessed} when redshift information from the DESI spectroscopic pipeline needs to be altered. This can occur when a DESI hardware glitch is identified after the MTL ledgers have already been updated for certain tiles, or due to improvements in the DESI spectroscopic pipeline software. Reprocessing of the ledgers is achieved by adding new entries to the ledger with the original state of each affected target, and then reprising the MTL updates, in the original tile-order, using the new redshift information. 

The root directory for the MTL ledgers includes two ``done'' files (named {\tt mtl-done-tiles.ecsv} and {\tt scnd-mtl-done-tiles.ecsv}) that list each tile that has been processed through the MTL logic.  These files communicate to afternoon planning that a tile's analysis is complete and overlapping tiles may be observed.  The files include a column (named {\tt ARCHIVEDATE}) that records when the redshift information used to update the MTL ledgers that touch a given tile was archived (\textsection\ref{subsec:archive}). As is the case for the other MTL ledgers, new entries are only ever appended to the ``done'' files (i.e.\ no information is ever overwritten). If a tile appears in a ``done'' file multiple times, then that tile was reprocessed, using information from redshifts on the recorded {\tt ARCHIVEDATE}. The corresponding ledgers will contain entries, in order, for both the original MTL state changes and any updates based on reprocessed redshift information.

\subsection{Other Ledgers}

Two bespoke types of MTL ledgers exist in addition to the five initial sets detailed in \S\ref{subsec:MTLinit}; a single, monolithic ledger listing targets of opportunity (henceforth ToO), and sets of ledgers used to override the MTL logic. 

The ToO ledger is read by {\tt fiberassign} to design special tiles to follow up gravitational wave detections, neutrino bursts, or other time-critical events \citep[e.g][]{Palmese:2021}. Entries in the ToO ledger can also be used to requisition fibers on existing tiles (see \S\ref{subsec:onthefly}), although this mode is yet to be used in the DESI Main Survey. The ToO ledgers differ from other MTL ledgers as they contain just the minimal information needed by {\tt fiberassign}, plus columns that are only relevant to time-critical observations.

Override ledgers are used to {\em force} an observational state {\em into} an MTL ledger. This is particularly beneficial when rare, high-value targets have been studied using newly available data and found to have a different redshift or classification to that assigned by the DESI pipeline. For example, the override mechanism currently ensures some quasars from a $z\sim5$ secondary program \citep{Yang:2023} --- which have been definitively classified through visual inspection of their DESI spectra --- are always available to receive a DESI fiber. Override ledgers closely resemble other MTL files, as they essentially contain the state that will be forced into an MTL ledger.

\section{Survey Performance}
\label{sec:performance}

Planning the DESI survey requires predicting the amount of effective time the survey can deliver over the year.  The amount of effective time delivered depends on the point spread function delivered to the focal plane (\textsection\ref{subsec:seeing}), the transparency of the night sky (\textsection\ref{subsec:transparency}), the sky brightness (\textsection\ref{subsec:sky}), the overall survey speed (\textsection\ref{subsec:overallspeed}), and the time off sky due to weather and technical downtime (\textsection\ref{subsec:downtime}).

In this and subsequent sections, we study the performance of the DESI survey from 2021--05--14 to 2022--06--14.  The start date corresponds to the start of the DESI main survey; after this point we limited engineering observations and observed almost exclusively main survey tiles.  The stop date corresponds to the beginning of a long shutdown due to damage to Kitt Peak infrastructure from the Contreras wildfire.  The DESI survey restarted operations on 2022--09--11; we do not include this more recent data here.

We compare DESI's performance with expectations from the Mayall Telescope's long history.  The Mayall has been observing the sky since 1973, providing a historical record of seeing, transparency, sky brightness, and downtime, based on the tireless, careful effort of the Mayall's observers.  We focus here particularly on the record from 2007--2017, where records were most readily available.  We compare DESI's observed performance with simulations based on on this historical record (\textsection\ref{sec:surveysim}).

An important concept in this section is the survey ``margin'': the amount of time available to the survey divided by the time needed to finish the survey, minus one.  DESI aims to operate with a healthy margin to enable finishing the survey in the allotted five year survey window.  Factors which speed the survey by a certain percentage increase the margin by the same percentage, in the limit that the margin is close to zero.

\subsection{Point spread function}
\label{subsec:seeing}
\begin{figure}
\includegraphics[width=\columnwidth]{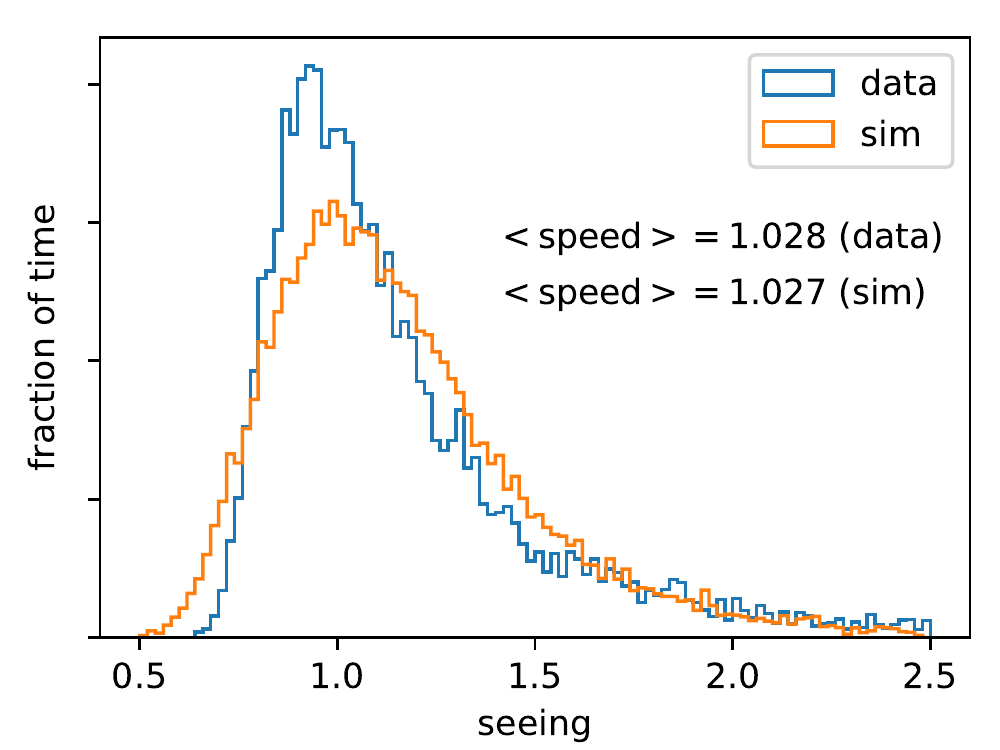}
\caption{DESI delivered point spread function FWHM.  The blue curve shows the measurements from the GFAs during the DESI survey, while the orange curve shows data from simulations based on the MzLS.  The inferred average survey speeds for both the real data and the simulated data (proportional to the square of the fraction of flux entering a fiber) is given for each case, and agree closely.  \label{fig:seeing}}
\end{figure}

The DESI corrector was designed to contribute negligibly to the PSF delivered to the focal plane.  This means that historical records from for example, the Mayall z-band Legacy Survey \citep[MzLS]{Dey:2019}, can be used directly to predict DESI's seeing.  Comparison of predictions from simulations (\textsection\ref{sec:surveysim}) and the actual seeing in the first year of the survey show good agreement, as shown in Figure~\ref{fig:seeing}.  The observed distribution is somewhat tighter than the simulated data based on the MzLS, plausibly due to DESI's improved control of focus using the GFAs.  However the overall inferred average speed (the square of the fraction of source flux entering a fiber, the critical element to survey planning) agrees closely with expectations from MzLS and the survey simulations.

\subsection{Transparency}
\label{subsec:transparency}
\begin{figure}
\includegraphics[width=\columnwidth]{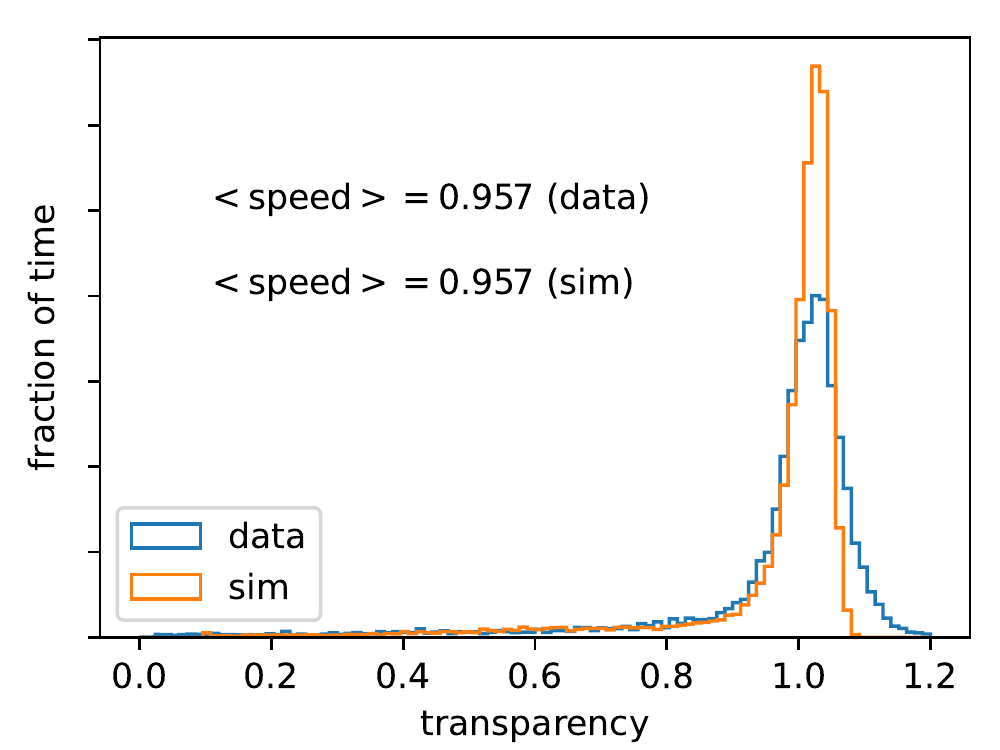}
\caption{DESI observed transparency.  The simulations show a narrower distribution of transparencies than observed, due to the simulations having deconvolved the observed distribution slightly to reduce the effect of measurement errors.  The inferred average survey speeds are proportional to the square of the transparency, and are identical between observations and simulations by construction.\label{fig:transparency}}
\end{figure}

Similarly, survey planning and simulations assume that the transparency of the night sky as seen by DESI will closely match the historical performance obtained by MzLS.  Again, predictions from simulations and DESI's observations in the first year show good agreement, as shown in Figure~\ref{fig:transparency}.  The average survey speed, proportional to the square of the transparency, shows excellent agreement between the data and the simulations, though this is by construction.

An unexpected challenge in matching the observations to the simulations stems from the definition of ``photometric.''  The distribution of transparencies seen by DESI (Figure~\ref{fig:transparency}) is strongly peaked near unity, but the peak has a width of about 3.5\%.  This width partially reflects measurement uncertainties, but also appears to reflect true variations in the transparency of the night sky, as confirmed by comparison with the amount of light delivered to the spectrographs and seen by the GFAs.  The nights that were used to define a transparency of 1 for DESI were $\sim 3\%$ less transparent than the peak of the transparency distribution.  The results shown in Figure~\ref{fig:transparency} have been updated to account for this discrepancy.

\subsection{Sky Brightness}
\label{subsec:sky}

\begin{figure}
\includegraphics[width=\columnwidth]{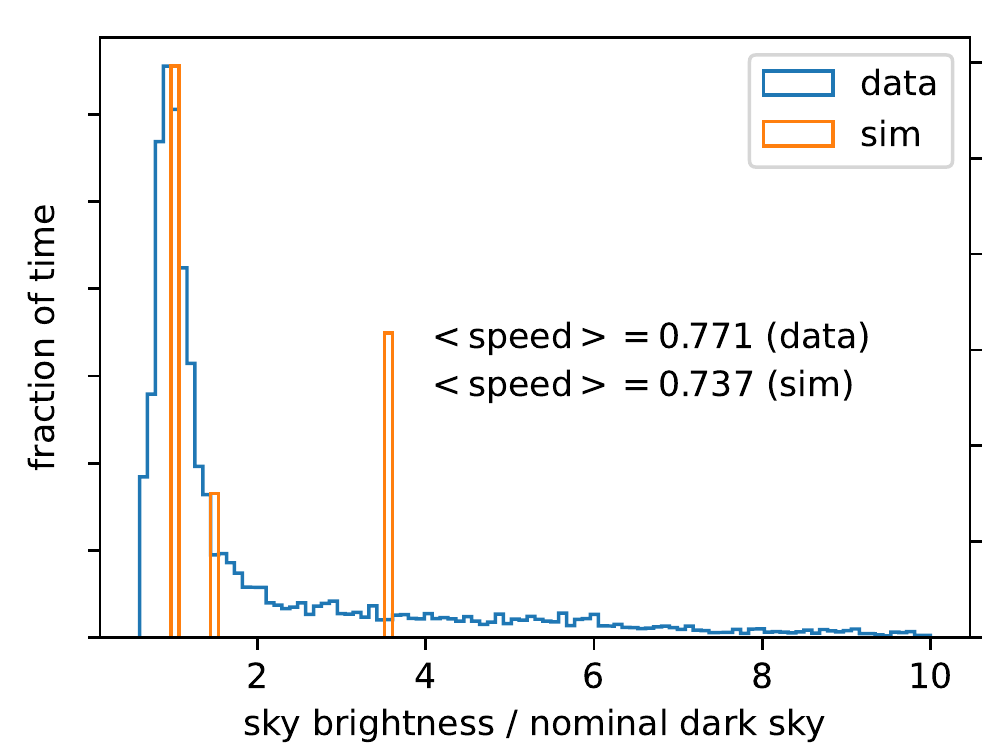}
\caption{DESI observed sky brightness, relative to a nominal dark sky brightness of 21.07 mag.  The observed sky brightness peaks about 7\% darker than this.  The sky brightness models in the simulation are very simple, assigning a sky brightness of 1, 1.5, or 3.6 depending on the phase and location of the moon.  The overall average survey speed, proportional to one over the sky flux, are reasonably well-matched, though the simulations are 8\% slower largely due to the slightly darker peak of the observed sky distribution than the simulated sky distribution.\label{fig:sky}}
\end{figure}

Survey planning focused on the main dark program, with less emphasis on the bright program, which accounts for only roughly 10\% of the survey effective time.  The sky brightness when the moon is up is a relatively complex function of the moon phase, location, and the line of sight.  However, when the moon is down, our model of the sky brightness is a simple function of airmass.  Survey planning then chose an extremely simple description of the sky brightness: equal to a nominal dark sky brightness when the moon is down; equal to $1.5\times$ nominal when the moon is up but less than 60\% illuminated and the product of the moon phase and distance from the horizon was smaller than 30 degrees; and equal to $3.6\times$ nominal otherwise.  Figure~\ref{fig:sky} compares this simple model in the simulations with DESI's observations.  The work of \citet{Hahn:2022} includes an improved sky model important for accurate modeling of the bright program.

This model is clearly limited, but because dark, moon-down time is the source of most of the survey's effective time, it is largely adequate.  The average survey speed, proportional to one over the sky flux, is about 8\% faster in the actual data than in the simulations.  This is largely because the dark sky brightness peaks 7\% darker than the nominal 21.07~mag forecast in survey planning.

\subsection{Overall speed}
\label{subsec:overallspeed}

\begin{figure}
\includegraphics[width=\columnwidth]{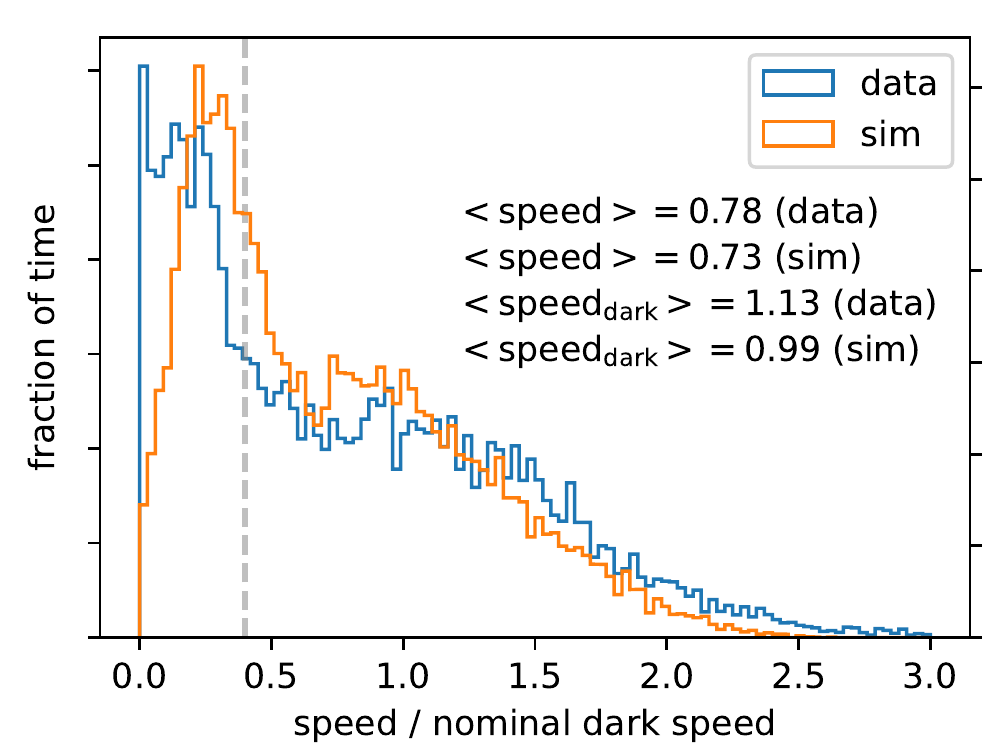}
\caption{DESI delivered survey speed, compared with speeds delivered in the simulations.  This is the product of factors relating to seeing, transparency, and sky brightness.  The average delivered speed is 7\% higher in the actual observations, but 14\% higher when limiting to exposures taken in dark conditions.  Note that the small difference between the average speeds here and in Figure~\ref{fig:surveyspeedhist} comes from the fact that here the speeds are computed from the measured seeing, transparency, and sky brightness, and in Figure~\ref{fig:surveyspeedhist} they are computed from the effective time delivered on each tile.\label{fig:speed}}
\end{figure}

The total delivered survey speed is a combination of the seeing, transparency, and sky brightness.  Breaking these terms out separately, one expects the simulations to run 8\% slower than the actual observations due to the different sky brightness modeling.  Figure~\ref{fig:speed} compares the actual total delivered speeds in the simulations and observations.

The observed average survey speed has been 7\% faster than expected in the simulations, consistent with the difference in sky brightness.  Additionally, the variance in the observed speeds is larger than predicted by our simple simulations, leading the average speed in the dark program---observed when conditions are good---to be 14\% larger than in the simulations.  This is the largest factor in leading to discrepancies between the observed and expected survey progress (see Table~\ref{tab:simobsmargin}).  This is largely driven by times when the skies are especially dark.

\subsubsection{Solar Cycles}
\label{subsubsec:solarcycles}

The DESI survey started survey validation near the start of solar cycle 25.  The next solar maximum will occur in 2025, near the end of the DESI main survey.  It is therefore likely that sky brightness distribution observed so far is darker than what we will have for the remainder of the survey \citep{Walker:1988, Patat:2008, Noll:2012}.  The impact of the solar cycle on DESI's overall performance will depend on the amplitude of the solar cycle.  Investigations using past data from the Sloan Digital Sky Survey and its extensions, as well as the DECam Legacy Survey \citep{Dey:2019}, suggested potential impacts on survey speed of between 5\% and 20\%.  For comparison, \citet{Patat:2008} measure an approximately 30\% difference in dark sky brightness between solar minimum and solar maximum.

\subsection{Downtime}
\label{subsec:downtime}

Another key element of survey planning is the amount of time the system is down, due to bad weather or technical problems with the system.  DESI's downtime has been very close to expectations, with the exception of two significant shutdowns in the summer of 2021 and 2022. Table~\ref{tab:downtime} lists the time lost to various causes, and the total time remaining.  We exclude the second shutdown from the time range considered in this work, but we describe it briefly here for completeness.

The first shutdown of the DESI main survey was from 2021--07--10 to 2021--09--20, when the focal plane electronics were upgraded.  The second shutdown was from 2022--06--14 to 2022--09--11, when a wildfire swept through Kitt Peak, requiring repair to the site's infrastructure.  Such large events are not directly incorporated into planning, and instead come out of the overall survey margin.  However, survey planning does include a nominal three week shutdown during Arizona's summer monsoon season, when nights are shortest and frequent clouds and rain slow observing.  Both shutdowns occurred during monsoon season, leading them to have a much smaller impact on survey progress than suggested by their duration.

Outside of these two shutdowns, DESI's downtime has been very modest and consistent with expectations.  The DESI performance database tracks the state of the system every second, recording a wealth of information, including whether the spectrograph shutter is open, whether the telescope is guiding, and whether the system is in a weather, instrument, telescope, or other hold.  Defining ``on sky'' time as time when the spectrograph shutter has been open while guiding within the last 2.5 minutes (to cover overheads between exposures and long slews), DESI has spent 76.6\% of its time on sky during ``dark time''.  Here we define ``dark time'' as time on nights more than two days from full moon, with the sun more than 15\degree\ below the horizon, with the moon down, and outside of one of the two major shutdowns.  The majority of the downtime (22.2\% of the dark time) is due to the weather, with another 1.4\% due to instrument downtime and less than 1\% to other sources.

\begin{deluxetable}{crr}[t]
\tablecaption{Dark Time Spent on Sky or Down\label{tab:downtime}}
\tablehead{\colhead{Category} & \colhead{\% of moon-down time} & \colhead{\% of all time}}
\startdata
On sky\tablenotemark{a} & 76.6\% & 69.1\% \\
Open shutter & 66.2\% & 58.4\% \\
Any recorded loss & 23.7\% & 30.3\% \\
Weather loss\tablenotemark{b} & 22.2\% & 27.9\% \\
Instrument loss & 1.4\% & 1.9\% \\
Telescope loss & 0.2\% & 0.4\% \\
Other loss & 0.4\% & 1.2\% \\
\enddata
\tablenotetext{}{Fraction of time spent either on sky or down, according to the DESI performance database.  We tabulate values for both ``moon-down'' time and ``all'' time.  ``All'' time includes all time outside of monsoon shutdowns with the sun more than 12\degree\ below the horizon.  ``Moon-down'' time is the subset of ``all'' time where the moon is below the horizon and excluding four nights around full moon.  Engineering activities take priority around full moon, and are concentrated in moon-up time in general, leading to better on-sky fractions in moon-down time.  The various sources of loss need not sum to the ``any recorded loss'', since the system can be down for multiple reasons simultaneously.  Small differences between 100\% and the sum of the on sky time and the any recorded loss time can stem from the definition of ``on sky'' time.
\tablenotetext{a}{On sky time is defined as time within 2.5 minutes of a moment when the spectrograph shutters were open and the telescope was guiding.}
\tablenotetext{b}{The weather loss here tabulates both time the observers mark as being lost due to weather as well as time when the instrument control system was not in ``observing'' mode.  The latter case usually corresponds to nights that cloud out but where the observers do not mark the time as a weather loss.  However, other, more rare cases will be incorrectly grouped with weather loss here.}
}
\end{deluxetable}

The instrument has has met the goal of $<2\%$ downtime, and other sources of downtime are negligible for planning purposes.  The weather loss of 22.2\% is typical for the Mayall outside of the major shutdowns DESI has experienced.  Specifically, replaying the years 2007--2017 as if they were 2021--05--14 to 2022--06--15, excluding time during major shutdowns, and weighting nights by the length of the night between 15\degree\ twilight, the Mayall would have been closed due to weather 23.7\% of the time on average, with a standard deviation of 3.6\%; DESI's observed weather loss so far of 22.2\% is typical.

The amount of time available for observation with the Mayall per month is given in Table~\ref{tab:hourspermonth}, based on the years 2007--2017.  This table uses the time between 15\degree\ twilight, adjusted for seasonal variability in the weather.  We have not removed planned engineering time around full moon or during the annual monsoon season, however, because the alignment of these shutdowns with month boundaries can artificially increase variability.

\begin{deluxetable}{crcr}[t]
\tablecaption{Weather-adjusted hours available per month\label{tab:hourspermonth}}
\tablehead{\colhead{Month} & \colhead{Hours} & \colhead{Month} & \colhead{Hours}}
\startdata
January      & $  240 \pm    47$ & July         & $  104 \pm    21$\\
February     & $  211 \pm    25$ & August       & $  148 \pm    26$\\
March        & $  240 \pm    21$ & September    & $  191 \pm    26$\\
April        & $  216 \pm    16$ & October      & $  258 \pm    36$\\
May          & $  201 \pm    14$ & November     & $  254 \pm    24$\\
June         & $  185 \pm    22$ & December     & $  222 \pm    27$\\
\hline
Annual & $  2468 \pm     80$ & &  \\
\enddata
\tablenotetext{}{The number of hours available for observation with the Mayall per month, accounting for varying weather and the changing length of the night, but excluding engineering and monsoon shutdowns.  Uncertainties reflect year-to-year standard deviations due to weather.}
\end{deluxetable}

As noted earlier, survey planning includes a three week shutdown around full moon during the Arizona monsoon season.  So far, our monsoon season shutdowns have been significantly longer than forecast there, owing to electronics upgrades and the Contreras wildfire.  On the other hand, we would plan to run DESI through the monsoon season if weather and engineering requirements allowed.  Table~\ref{tab:hourspermonth} gives a sense for how much that adjustment would speed the survey---recovering the bright part of July would be roughly a quarter as valuable as a January.

\subsection{Effective hours delivered per year}
\label{subsec:totaleffectivetime}
When planning programs for DESI, it can be valuable to have a sense for the total number of effective hours DESI can deliver in a year.  Table~\ref{tab:totefftime} tabulates some key numbers for making this calculation.

\begin{deluxetable}{lrl}[t]
\tablecaption{Amount of Effective Time per Year\label{tab:totefftime}}
\tablehead{\colhead{Parameter} & \colhead{Value} & \colhead{Notes}}
\startdata
Time per year\tablenotemark{a} & 3481 hr & Planning \\
Open shutter fraction & 58.4\% & observed \\
Fraction of dark time\tablenotemark{c} & 59.3\% & observed \\
Fraction of bright time\tablenotemark{c} & 34.8\% & observed \\
Fraction of backup time\tablenotemark{b,c} & 5.9\% & observed \\
Average dark speed & 1.148 & observed \\
Average bright speed & 0.293 & observed \\
Average backup speed\tablenotemark{b} & 0.096 & observed \\
Average overall speed & 0.789 & observed \\
Dark effective time per year & 1383 hr & computed \\
Bright effective time per year & 207 hr & computed \\
Backup effective time per year\tablenotemark{b} & 12 hr & computed \\
\hline
Number of dark tiles & 9929 & design \\
Number of bright tiles & 5676 & design \\
Effective time for dark tiles & 1000 s & design \\
Effective time for bright tiles & 180 s & design \\
Effective time for backup tiles & 60 s & design \\
Mean airmass \& dust adjustment & 1.51 & design \\
Dark time needed per year & 833 hr & computed \\
Bright time needed per year & 86 hr & computed \\
\hline
Outside major unplanned shutdowns\tablenotemark{d} & 87\% & observed \\
Time on tiles not counted\tablenotemark{e} & 2\% & observed\\
Average dark tile overexposure\tablenotemark{f} & 2\% & observed\\
\hline
Dark margin, computed & 39\% & computed \\
Bright margin, computed & 105\% & computed \\
Dark margin, observed & 36\% & observed \\
Bright margin, observed & 93\% & observed \\
\hline
\enddata
\tablenotetext{}{Parameters controlling the amount of effective time available to the survey (top of table), compared with parameters controlling the time needed to complete the survey (middle of table).}
\tablenotetext{a}{The number of hours derived from ephemerides; see \textsection~\ref{subsec:totaleffectivetime} for details.}
\tablenotetext{b}{Backup program parameters are especially uncertain because backup tiles were not regularly observed until December 2021.}
\tablenotetext{c}{We are defining the time available to the program according to the amount of time selected for that program based on the NTS program selection.  See \textsection\ref{subsec:speed} for more details.}
\tablenotetext{d}{This fraction is the expected time available to the survey given the long summer 2021 shutdown divided by what the survey would have had with the planned shutdown.}
\tablenotetext{e}{Tiles ``not counted'' as main survey tiles were either observed for other programs (1\%) or discarded (1\%).
\tablenotetext{f}{The average completed dark tile has 1.02$\times$ the required effective time.}
}
\end{deluxetable}

We were able to get a good match between the observed dark margin and the margin expected from a relatively simple calculation based on the number of hours available to the survey and the survey's average speed in different programs.  The calculations count every hour with the sun more than 12\degree\ below the horizon, excluding an 18 night shutdown around full moon each monsoon season for engineering purposes.

Matching the computed margin to the actual margin requires accounting for the longer-than-expected DESI shutdown in the summer of 2021 (\textsection\ref{subsec:downtime}).  Other small adjustments are needed to account for time DESI has spent on tiles for programs other than the main survey (1\%) and on exposures that needed to be discarded (e.g., due to wind shake, or temporary instrument problems; 1\%).  

Note that this calculation folds in true values of critical parameters DESI achieved during the 2021--05--14 to 2022--06--15 time window under consideration---it uses the observed open shutter fraction and the observed average speeds and fractions of time in different programs.  This effectively folds in the real weather and conditions that DESI has experienced and all technical downtime.  These values are useful for the planning of future DESI-like surveys, but the match between the observed DESI margin and the computed value from this computation is somewhat artificial.

We can check the consistency of this table by comparing the number of hours accumulated on dark tiles between 2021--05--14 and 2022--06--15 with the expectations from this table.  On the basis of the ephemerides, there are 3248 total hours excluding the long shutdown in the summer of 2021.  Using the open shutter fraction, fraction of time in the dark program, and average dark program speed from Table~\ref{tab:totefftime}, we obtain 1291 effective hours at zenith through no extinction.  Counting all time accumulated on dark exposures in that window, and adjusting by Equation~\ref{eq:dust} and Equation~\ref{eq:airmass} to account for extinction and airmass, we obtain 1247 observed effective dark hours.  These are different by 3.5\%.  Much of the difference is ``time on tiles not counted'', e.g., time we spent observing tiles for special programs or tiles that we eventually deemed bad.  Another issue surrounds the accounting for engineering time; engineering time spent on guided observations with the spectrographs open counts as open shutter time in Table~\ref{tab:totefftime}, though this kind of open shutter time needs to be separately accounted when computing the amount of time DESI can deliver on science tiles.  Still, these are small effects, and Table~\ref{tab:totefftime} provides a useful description of the number of effective hours the DESI system can deliver.

\section{Survey Simulations}
\label{sec:surveysim}

We perform survey simulations to verify that the DESI survey will complete in its allotted five-year mission.  The survey simulations step through the survey at ten second intervals each night of observations.  The simulation generates a realistic realization of the observing conditions (seeing, transparency, sky brightness) based on modeling of past observing conditions from the Mosaic z-band Legacy Survey \citep[MzLS]{Dey:2019}.  Downtime due to weather is also included, following patterns from observations at the Mayall from 2007--2017.

At each time step, if the system is not already observing, a new tile is selected, and the telescope begins tracking a new field overhead (Table~\ref{tab:simparams}).  Otherwise, when the system is observing, effective time is accumulated according to the current seeing, sky brightness, and transparency.  Observing continues until the tile is complete or the tile needs to be split or abandoned due to overly long exposures or too-high airmass.  When splitting, a separate tile split overhead is incurred (Table~\ref{tab:simparams}).  Weather-related downtime may also close the dome at any point, stopping the current observation and advancing the simulation to the next time the dome opens.

\begin{deluxetable}{cr}[t]
\tablecaption{Selected Survey Simulation Parameters\label{tab:simparams}}
\tablehead{\colhead{Parameter} & \colhead{Value}}
\startdata
Nightly beginning \& end of observations & 15\degree\ twilight \\
New field overhead & 139~s \\
Split exposure overhead & 70~s \\
Engineering nights per lunation & 4 \\
Monsoon shutdown nights per year & 18 \\
\enddata
\tablenotetext{}{A selection of important parameters in the simulations, and their values.}
\end{deluxetable}

The survey simulations use the same airmass optimization and next-tile selection algorithms as the real survey.  Accordingly the simulations follow the same moon \& planet avoidance algorithms as the real survey.  They use a simplified model of the ETC and a simple model of the instrument.  They model only per-tile quantities and ignore any details relating to individual fibers and target selection; the survey simulations seek only to accumulate the required effective time on each tile.

The survey simulations include realistic models of the weather based on historical data from the Mayall.  Comparisons of modeled seeing, transparency, sky brightness, and delivered speed are shown in Figures \ref{fig:seeing}, \ref{fig:transparency}, \ref{fig:sky} and \ref{fig:speed}.  The sky modeling in the simulations is rudimentary, but the seeing and transparency distributions match the observations closely.  Moreover, the time correlation of the variations in the seeing and transparency is modeled with a Gaussian process, with power spectral densities chosen to closely match observations from the MzLS.  That said, the accuracy of the time correlations of variations in the weather makes only a minor impact on survey planning.

Overheads due to stopping and splitting exposures are modest.  For the dark program as of 2022--10--04, the mean exposure time is 1093~s, over 3725 observations of 2913 tiles.  This implies an overhead of about 9\%, which is well captured by the simulations.  Slew time is ignored in the simulations, and would account for an additional overhead of about 3\%, using the slews from a simulated survey and a realistic model for the telescope slew time as a function of the change in hour angle and declination.

The survey simulations can incorporate past data and use them to make forecasts for the future given different scenarios.  This is valuable to, for example, understand the impact of different planned maintenance activities requiring shutting down the telescope to the final survey margin.

\subsection{Comparing survey simulations with the observed survey progress}
\label{subsec:simobs}

\begin{figure*}
\includegraphics[width=\textwidth]{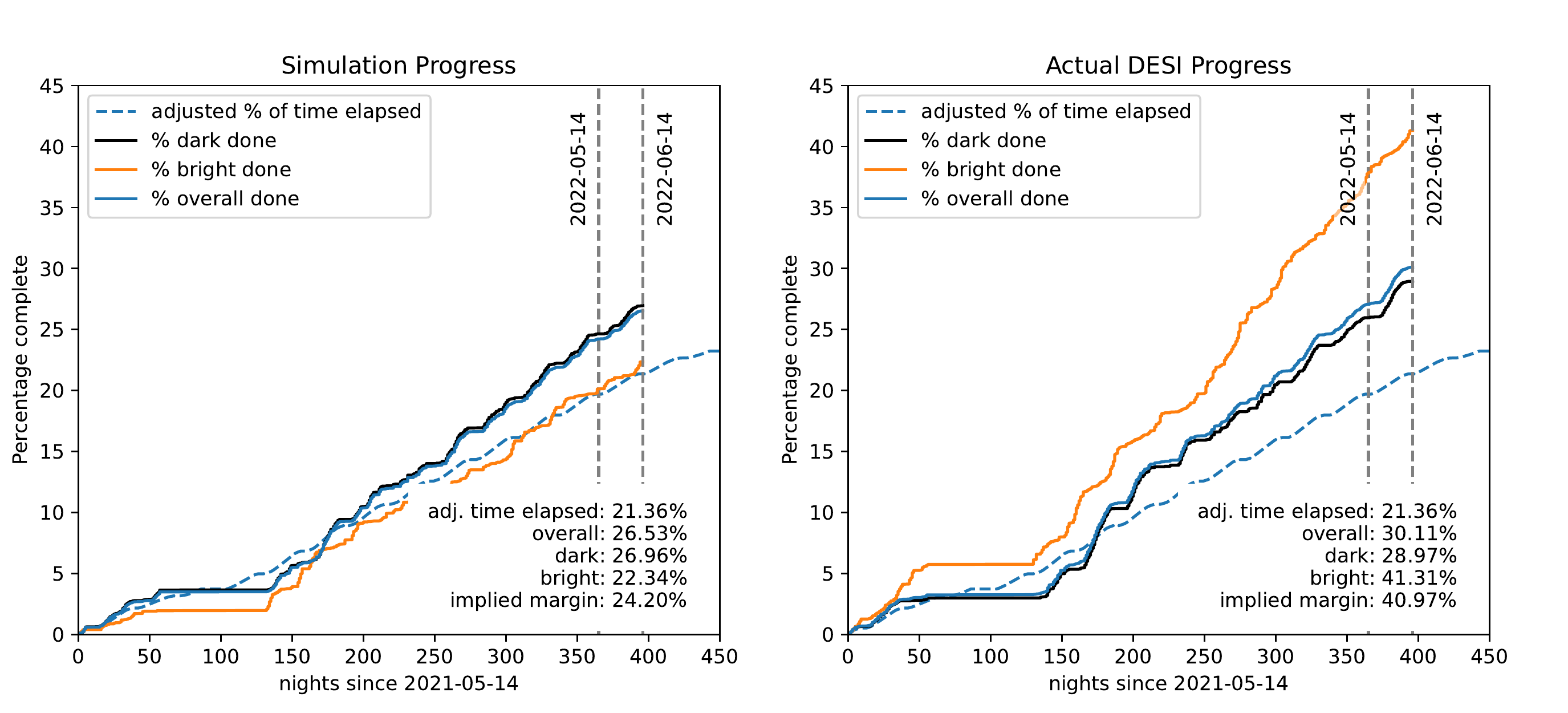}
\caption{DESI observed progress compared with a nominal simulation using the same major shutdowns.  The dark time progress of the simulation is a good match for the observed dark time progress; in the simulation, 26.53\% of the dark program was completed before 2022-06-15, while in the real survey, 28.97\% of the survey was completed.  The fraction of time elapsed is shown with a dashed line, weighting nights by the length of the night, historical weather loss, and removing nights near full moon and planned monsoon shutdowns; see \textsection\ref{tab:totefftime} for details.  These survey simulations match the progress of the bright program poorly, however, with the actual bright survey progress running ahead of the simulations by almost a factor of two.  This is due to limitations of the sky brightness modeling in the simulations, as well as the use of more time in twilight and near full moon for bright observations than expected.\label{fig:surveysim}}
\end{figure*}

Figure~\ref{fig:surveysim} shows an example survey simulation run.  For this run, we chose to exactly duplicate DESI long summer 2021 shutdown, as described in \textsection\ref{subsec:downtime}.  No additional sources of downtime were included except for normal weather losses, which were chosen to replicate randomly-sampled years of the Mayall's historical weather record.

The survey simulation matches the dark program reasonably well.  In the survey simulation, 26.96\% of the dark program is completed before 2022-09-21, while in the real survey, 28.97\% of the survey was completed.  The DESI survey is proceeding 7\% faster than forecast in the simulations, our top line result.  However, the comparison is complicated by the different average speed in the dark program in the simulations than in reality; see \textsection\ref{subsec:overallspeed}.  Accounting for this makes the dark program 14\% faster while being the active program on the telescope for 3\% less time than expected.  Additional minor differences between the simulations and real observations are that the simulations neglect slew overheads and technical downtime (3\% and 2\% effects).  More importantly, the simulation year one weather realization is particularly poor, with 11\% more lost time than DESI observed from 2022--05--14 to 2023--06--15, outside the summer 2021 shutdown.  Finally, 2\% of the time in the real survey was spent either on tiles we end up discarding or on tiles that were not for the main survey, and another 2\% of time was spent overexposing dark tiles.  Table~\ref{tab:simobsmargin} summarizes the different contributions to discrepancies between the simulation completeness and the observed completeness.  We conclude that the main survey is running 4\% slower than we would expect from the simulations after accounting for all of these effects, which we consider good agreement.

\begin{deluxetable}{lr}[t]
\tablecaption{Contributors to differences in dark margin \label{tab:simobsmargin}}
\tablehead{\colhead{Cause} & \colhead{Fraction}}
\startdata
Observed progress through 2022-06-14 & 29.0\% \\
Simulated progress through 2022-06-14 & 27.0\% \\
Expected effective time through 2022-06-14 & 21.4\% \\
\hline
Dark speed & +14\% \\
Fraction of time in dark program & $-3\%$ \\ 
Neglected slew time & $-3\%$ \\
Neglected technical downtime & $-2\%$ \\
Actual weather versus simulated & $+11\%$ \\
Time on tiles not counted & $-2\%$ \\
Dark tiles are overexposed & $-2\%$ \\
\hline
Adjusted simulated completeness & 30.2\%\\
Ratio of observed and simulated completeness & $+7\%$ \\
Ratio of completeness after adjustments & $-4\%$ \\
\enddata
\tablenotetext{}{Important contributions to the difference between the observed completeness in the simulations and the actual observed completeness of the survey.  The signs are chosen so that improving the simulations would change the simulated completeness in the indicated direction.  A number of minor effects are present, which together would lead the simulations to run 12\% faster, exceeding the 7\% difference between the observed and simulated completeness.  A large number of effects come into play.}
\end{deluxetable}

We have focused on the dark program, which accounts for most of DESI's effective time, and for which the survey simulations are best suited.  The bright program is running much faster than expected from the simulations, due primarily to the following:
\begin{itemize}
    \item The simulations include no observations when the sun is within 15\degree\ of the horizon; in fact we aim to start observing the backup program at 10\degree\ twilight and the bright program at 12\degree\ twilight.
    \item The simulations include no observations within 4 days of full moon; in practice, this time is often used for observing when no engineering work is planned.
    \item The simulation sky modeling in bright conditions is rudimentary. (\textsection\ref{subsec:sky}).
\end{itemize}
The bright program was more than 40\% complete prior to the summer 2022 shutdown, after little more than a year of main survey observations!  This program will need to be expanded in order to accommodate the available time.

\section{Conclusion}
\label{sec:conclusion}

The Dark Energy Spectroscopic Instrument's main survey began on 2021--05--14, and has observed more than 14~million galaxies and 4~million stars through 2022--06--14.  The success of the survey has relied on the efforts and dedication of a large science collaboration, instrument, and operations team.  The DESI instrument's performance largely exceeds expectations; the data management, processing, and analysis routinely delivers high quality redshifts within hours of observation, even while accommodating last-minute changes in instrument configuration \& calibrations; and the operations team has put together a robust system to feed back past observations into the design of future observations on a daily basis, while identifying and removing problematic observations.  The collaboration's realization of the scientific potential of these observations is now underway.

We have laid out the choices made in the survey strategy---the survey footprint, the amount of observing time needed on each tile, the hour angles at which the tiles should be observed, and the tiles' priorities.  The decision to require that all observations be fully processed before making subsequent overlapping observations allows the survey to reobserve any $z>2.1$ quasar discoveries, and places strict requirements on the daily operations design and plan.  We detailed the steps of the daily operations loop largely implied by this decision, from afternoon planning to nightly observations to data reduction to updating DESI Merged Target Lists.  These Merged Target Lists play a central role in tracking DESI observations in operations, and we described the details of their construction and updates following targets' observation.

We also described the survey performance, which has somewhat exceeded projections made on the basis of historical data from the MzLS---the sky has been slightly darker than we expected.  Instrument downtime has been kept low (excepting a major shutdown during the summer monsoon season for upgrading the focal plane electronics), leaving the survey with a healthy 36\% margin on 2022--06--14.  We compared the observed survey performance with detailed simulations and found good agreement, increasing our confidence in the simulations' value for predicting survey performance.

The first 1.1 years of DESI's operations have been an exciting success, and we look forward to a long, productive future for the instrument.

\acknowledgements

ADM was supported by the U.S.\ Department of Energy, Office of Science, Office of High Energy Physics, under Award Number DE-SC0019022.

This material is based upon work supported by the U.S. Department of Energy (DOE), Office of Science, Office of High-Energy Physics, under Contract No. DE–AC02–05CH11231, and by the National Energy Research Scientific Computing Center, a DOE Office of Science User Facility under the same contract. Additional support for DESI was provided by the U.S. National Science Foundation (NSF), Division of Astronomical Sciences under Contract No. AST-0950945 to the NSF’s National Optical-Infrared Astronomy Research Laboratory; the Science and Technologies Facilities Council of the United Kingdom; the Gordon and Betty Moore Foundation; the Heising-Simons Foundation; the French Alternative Energies and Atomic Energy Commission (CEA); the National Council of Science and Technology of Mexico (CONACYT); the Ministry of Science and Innovation of Spain (MICINN), and by the DESI Member Institutions: \url{https://www.desi.lbl.gov/collaborating-institutions}. Any opinions, findings, and conclusions or recommendations expressed in this material are those of the author(s) and do not necessarily reflect the views of the U. S. National Science Foundation, the U. S. Department of Energy, or any of the listed funding agencies.

The authors are honored to be permitted to conduct scientific research on Iolkam Du’ag (Kitt Peak), a mountain with particular significance to the Tohono O’odham Nation.

\vspace{5mm}
\facilities{DESI}
\software{astropy \citep{astropy2013a, astropy:2018, astropy:2022}}

\appendix

\section{Airmass Optimization}
\label{sec:airmassapp}

The DESI airmass optimization scheme works by assigning local sidereal times to tiles and computing the total time necessary to observe the tiles given that assignment.  It aims to minimize a cost $C$:
\begin{align}
    C & = T + R \\
    T & = \frac{T_p - T_0}{T_0} \label{eq:totsurveytime} \\
    R & = \frac{1}{\sum_i^n{P_i}}(n\sum{(sA_i - P_i)^2})^{1/2} \label{eq:rmse} \\
    s & = \sum P_i / \sum A_i \, ,
\end{align}
where $T_p$ is the total time needed to observe the survey given the planned local sidereal times and the implied airmasses, and $T_0$ is the time that would be needed to observe the survey were all tiles observed at an hour angle of 0.  $P_i$ and $A_i$ are the number of planned and available hours in a particular bin $i$ of LST, and $n$ is the total number of bins of LST used.  Note that hour angles $HA$ and LSTs are related by $HA = LST - \alpha$, and that assigning an LST to a tile is equivalent to assigning an hour angle to a tile, since each tile has a defined right ascension $\alpha$.

More explicitly, the total times $T_P$ and $T_0$ are given by
\begin{align}
    T_0 & = \sum T_{0, i} \\
    T_P & = \sum T_{H, i} \\
    T_{H,i} & = G_i \, 10^{2 \times 2.165 \times E(B-V) / 2.5} X_{i, H}^{1.75} \,
\end{align}
where $T_{H, i}$ is the estimated time needed to observe tile $i$ at an hour angle of H, $X_{i, H}$ is the airmass of tile $i$ at hour angle $H$, and $G_i$ is the goal time for a tile (1000~s for a dark tile or 180~s for a bright tile).  Note that sky brightness variations due to the moon are not accounted for here, and that one obtains the same solution for any $G$ as long as it is constant in a program, as for DESI.

The term $T$ (Equation~\ref{eq:totsurveytime}) is proportional to the total observing time (up to an additive constant); we want to minimize it.  The term $R$ (Equation~\ref{eq:rmse}) is the root mean square difference between the binned, planned LST distribution and the available LST distribution.  It is zero if the distribution of LST available to the survey exactly matches the planned distribution of LST.  An alternative optimization algorithm would force these two quantities to match; the approach taken here allows these to diverge but includes the divergence in the cost function $C$.  For DESI we choose bins 1.875\degree\ in size when binning the available and planned LST distributions $A_i$ and $P_i$.  

Our approach to assigning LSTs to tiles starts with an initial guess.  This initial guess is then optimized by a simulated annealing algorithm, which perturbs the assignment to try to reduce the cost $C$.

To create the initial hour angle assignments, we first construct the cumulative distribution function of the tiles' observational costs as a function of right ascension, $CDF_O(\alpha)$.  To construct this, we need to know what the observational cost of a tile is, and for that we need the tile's airmass---but we do not know the tile's airmass because we have not yet assigned it an hour angle.  For this initial guess we presume that all tiles will be observed with an hour angle of zero.  We also construct the cumulative distribution function of the available LST, $CDF_L(L)$, choosing $CDF_L(L_\mathrm{start}) = 0$ and integrating around the circle.  We then find for each right ascension $\alpha_i$ the corresponding LST $L_i$ such that $CDF_O(\alpha_i) = CDF_L(L_i)$.  Conceptually, this corresponds to matching the first 10\% of the tiles in right ascension to the first 10\% of the LSTs (starting from $L_\mathrm{start}$), and so on, until all tiles have been mapped to LSTs.  This gives a mapping of tiles to LST that provides the initial guess for the simulated annealing.  The only free parameter in this initial guess is $L_\mathrm{start}$, the LST at which to start the cumulative distribution function; this corresponds to the LST to which to map tiles with $\alpha = 0\degree$.  We choose a number of $L_\mathrm{start}$ values around the unit circle and use the $L_\mathrm{start}$ with the best score to produce the initial guess.

The simulated annealing process consists of a number of steps. In each step, we start by identifying LSTs where changing the assignment of LSTs to tiles by one bin in LST would most significantly improve $R$, the component of the cost coming from the difference between the planned and available times.  These bins are identified by finding the locations where $|\Delta(sA_i - P_i)|$ is largest, where $\Delta$ represents taking the difference between bin $i$ and bin $i - 1$.  One of the top five such bins is selected at random.  A scale factor is chosen from a Rayleigh distribution.  The LST of each tile $j$ in the selected bin is adjusted by the scale factor and the new survey cost $C_j$ is computed.  The new plan with the minimum $C_j$ is chosen (if any is better than the original $C$), and the process repeats.  If instead no improvement was found, instead 20\% of tiles are selected at random.  Then again the LST assignment of each of these tiles is perturbed, the new cost $C$ is computed, and the assignment with the best $C$ is kept.

The simulated annealing steps are grouped into rounds.  Each round consists of one simulated annealing step per tile in the program being optimized (i.e., 9929 steps for the dark program, and 5676 steps for the bright program).  When a round is complete, the LST assignment to tiles is mildly smoothed. 
 Each tile's hour angle is replaced by $H_i^\prime = (1 - \alpha) H_i + \alpha \bar H_i$, where $\bar H_i$ is the hour angle map convolved with a Gaussian with a length of 10\degree, and $\alpha$ is a parameter between 0 and 1 reflecting how aggressively to replace the hour angles with the smoothed version.  This smoothing is expected to improve the cost, because the optimal solution should assign LSTs to tiles in a spatially smooth manner.  Next, the perturbation scale is reduced to 95\% of its previous value, from an initial values of 1\degree.  Finally, $\alpha$ is reduced to 95\% of its previous value, from an initial value of 5\%.  Then another round of simulated annealing is performed with the updated parameters.  Rounds continue until both $R < 0.02$ and the fractional improvement in $C$ is less than 1\%, $C_i / C_{i-1} - 1 > -0.01$, where $i$ indexes rounds.

In practice, the simulated annealing scheme does not shift the solution far from the initial guess.  The primary limitation of the initial guess is that it gives all of the tiles at the same right ascension the same LST.  An optimal solution, however, keeps tiles at low declination close to hour angles of zero and preferentially uses tiles at high declination to fill in the LST distribution.  Experiments with alternative optimization schemes only improved the cost by roughly half of one percent.

\bibliography{surveyops}
\end{document}